\documentstyle[12pt,twoside,draft]{article}
\newif\iffigures
\iffigures
\fi

\newif\ifamsfonts
\amsfontstrue
\ifamsfonts
\font\twlbbb=msbm10 scaled\magstep1
\font\egtbbb=msbm8
\font\sixbbb=msbm6
\newfam\bbbfam
\textfont\bbbfam=\twlbbb
\scriptfont\bbbfam=\egtbbb
\scriptscriptfont\bbbfam=\sixbbb
\newcommand{\Bbb}[1]{{\fam\bbbfam\relax#1}}
\else
\newcommand{\Bbb}[1]{{\bf#1}}
\fi

\def\Hset{\Bbb{H}}
\def\Cset{\Bbb{C}}

\def\Rset{\Bbb{R}}
\def\Sset{\Bbb{S}}
\def\Tset{\Bbb{T}}

\def\Zset{\Bbb{Z}}
\def\Gset{\Bbb{G}}

\newtheorem{lem}{Lemma}[section]

\newtheorem{thm}{Theorem}[section]
\newtheorem{prop}{Proposition}[section]

\newtheorem{dfn}{Definition}[section]

\newcommand{\proof}{\noindent{\it Proof. }\ignorespaces}
\newcommand{\qed}{\relax\hfill\mbox{$\Box$}\par\vskip\topsep}

\renewcommand{\theequation}{\thesection.\arabic{equation}}
\makeatletter\@addtoreset{equation}{section}\makeatother

\newcommand{\sn}{\mathop{\rm sn}\nolimits}

\newcommand{\sech}{\mathop{\rm sech}\nolimits}

\newcommand{\trace}{\mathop{\rm trace}\nolimits}
\newcommand{\abs}[1]{\left | #1 \right |}

\newcommand{\Or}{\mathop{\rm O}\nolimits}

\newcommand{\id}{\,{\rm d}}
\newcommand{\un}{{\rm u}}
\newcommand{\st}{{\rm s}}

\newcommand{\ce}{{\rm c}}

\newcommand{\iu}{\mskip2mu{\rm i}\mskip1mu}

\def\w{\cal}
\def\rh{\rightarrow}
\def\lgh{\longrightarrow}

\def\pt{\partial}
\def\bd{\begin{displaymath}}
\def\ed{\end{displaymath}}
\def\bqns{\begin{eqnarray*}}
\def\bi{\begin{itemize}}
\def\ei{\end{itemize}}
\def\beq{\begin{quote}}
\def\eeq{\end{quote}}
\def\ben{\begin{enumerate}}
\def\een{\end{enumerate}}
\def\eqns{\end{eqnarray*}}
\def\bq{\begin{equation}}
\def\bqn{\begin{eqnarray}}
\def\eq{\end{equation}}
\def\eqn{\end{eqnarray}}

\def\a{\alpha}
\def\b{\beta}
\def\e{\varepsilon}
\def\l{\lambda}

\def\m{\mu}
\def\mm{\rm m}
\def\th{\theta}

\def\G{\Gamma}
\def\hr{\tilde r}
\def\k{\rm k}

\def\d{\delta}
\def\D{\Delta}
\def\o{\omega}
\def\s{\sigma}
\def\S{\Sigma}
\def\O{\Omega}

\def\k{\kappa}

\def\rA{\rm A}

\def\rB{\rm B}
\def\rC{\rm C}

\def\rL{\rm L}
\def\rE{\rm E}
\def\rK{\rm K}
\def\rG{\rm G}

\def\rP{\rm P}

\def\rx{\rm x}
\def\rF{\rm F}
\def\rH{\rm H}

\def\di{\displaystyle}

\renewcommand{\theequation}{\thesection.\arabic{equation}}

\newcounter{saveeqn}
\newcommand{\alpheqn}{\setcounter{saveeqn}{\value{equation}}% 
\stepcounter{saveeqn}\setcounter{equation}{0}% 
\renewcommand{\theequation}{\mbox{\thesection.\arabic{saveeqn}\alph{equation}}}}
\newcommand{\reseteqn}{\setcounter{equation}{\value{saveeqn}}%
\renewcommand{\theequation}{\thesection.\arabic{equation}}}

\begin{document}                                       
\setcounter{page}{1}
%\pagestyle{myheadings}
%\markboth{V.M. Rothos}{Mel'nikov Analysis and perturbed sine-Gordon equation}

\title{{\bf Mel'nikov Analysis of Homoclinic Chaos in a Perturbed
sine\,-\,Gordon Equation}
}
\author{ Vassilis M. Rothos{\footnote{E-mail: V.Rothos@damtp.cam.ac.uk}}\\
The Nonlinear Centre\\
Department of Applied Mathematics and Theoretical
Physics\\ University of Cambridge\\
 Silver St., Cambridge
CB3 9EW, U.K
}
\date{}
\maketitle
\begin{abstract}
We describe and characterize rigorously the chaotic behavior 
of the sine--Gordon equation. The existence of invariant manifolds and the
persistence of homoclinic orbits for a perturbed sine--Gordon equation
are established. We apply a geometric method based on Mel'nikov's analysis 
to derive conditions for the
transversal intersection of invariant manifolds of a hyperbolic point of the
perturbed Poincar\'e map. 
\end{abstract}
%\vskip 1.5 true cm
%\tableofcontents
%\addcontentsline{toc}{appendix}{Appendix}
%\addtocontents{toc}{Appendix}
%\newpage
\section{Introduction}
\setcounter{equation}{0}
Homoclinic orbits have long been identif\/ied as a possible source of chaos in
nonlinear f\/inite--dimensional dynamical systems. For example, when an
integrable 
2-dimensional system, containing a smooth connection between the invariant
manifolds of a saddle f\/ixed point is perturbed, this connection is generally
broken and the invariant manifolds intersect transversally. This yields
isolated homoclinic orbits of the perturbed system, near which one expects to
find shift map embeddings and chaos.

In the theory of f\/inite--dimensional dynamical systems, a standard method to
convert numerical experiments of chaotic dynamics into a rigourous mathematical
description is to construct Smale horseshoes in a neighborhood of a homoclinic
orbit \cite{wig1} and then use these horseshoes to establish the existence of
an invariant set on which the motion is topologically equivalent to a Bernoulli
shift on a number of symbols. Such constructions begin with the existence,
persistence and breaking of homoclinic orbits, which in turn are obtained
through Mel'nikov analysis. 

However, in the case of inf\/inite dimensional dynamical systems the situation is
more complicated \cite{tem}. As is well--known, realistic physical systems are
usually modelled by nonlinear PDEs, whose chaotic behaviour is generally very 
dif\/f\/icult to study analytically. In the last 10 years, 
D.W.McLaughlin and collaborators have studied the periodic spectral transform of
integrable soliton systems and constructed representations of global objects
such as homoclinic orbits and whiskered tori, extending the Mel'nikov
analysis to the NLS equation \cite{li}, \cite{dmac}-\cite{dmac3}, as
well as a system of coupled NLS \cite{r5}, \cite{r6}. Recently,
G. Haller has been developing an alternative prespective, under
certain conditions, he proved the existence of manifolds of
multi-pulse Silnikov type orbits homoclinic to the critical torus of
NLS equation. He has applied to f\/inite dimensional discretizations of
the perturbed NLS equation \cite{halwig}, \cite{hal98} and to the PDE
case \cite{hal99a}, \cite{hal99b}.

In this article we describe and characterize with mathematical precision
the chaotic behavior of the perturbed sine--Gordon equation (SG)
\alpheqn
\bq
{\rm Case}\,{\rm I}\quad u_{tt}-c^2u_{xx}-{\sin}\,u={\e}(g_{1}(u)-au_{t})
\label{1.2a}
\eq
\bq
{\rm Case}\,{\rm II}\quad u_{tt}-c^2u_{xx}-{\sin}\,u={\e}(g_{2}(t)-au_{t})
\label{1.2b}
\eq
\reseteqn
with even periodic spatially boundary conditions, $\e$ a small perturbation
parameter, $0< {\e} \ll 1$. The perturbation term $g_{1}(u)$ is a
smooth nonlinear function satisfying $g_{1}(0)=0, {\pt}_{u}g_{1}(0)=0$ and
$g_{2}$ is time-periodic with period $T$.
 
The unperturbed sine--Gordon equation  
\bq
u_{tt}-c^2u_{xx}-{\sin}\,u=0
\label{1.2}
\eq
is an inf\/inite--dimensional Hamiltonian
system \cite{tem}.
We denote by ${\Hset}^{1}$ the Sobolev space of functions of $x$ which are 
$L$--periodic, even and square integrable with square integrable first
derivative on $[0, L)$
\bq
{\Hset}^{1}=\Big\{\,{\bf u}=(u, u_{t})\,:\,{\bf u}(-x)=
{\bf u}(x)={\bf u}(x+L),\quad 
\int_{0}^{L}{\vert {\pt}_{x}{\bf
u}\vert}^{2}{\id}x<{\infty}\,\Big\}\equiv {\w X}
\label{space}
\eq
The system of sine-Gordon 
models the phase dif\/ference between two superconducting layers in a Josephson
junction, \cite{ped}.
Equation (\ref{1.2}) can also be thought of as a model for the continuum limit
of a 
chain of coupled pendula, including damping and driving, with
$u(x, t)$ the angle from the vertical down position of the pendulum at
the point $x$ along the chain.

On the phase space ${\Hset}^{1}$, the Hamiltonian of the equation (\ref{1.2})
is given by
$$H: {\Hset}^{1}\lgh {\Rset},\quad 
H({\bf u})=\int_{0}^{L}\left(\frac{1}{2}(u^{2}_{t}+c^2u^2_{x})+ 
(1+{\cos}\,u)\right){\id}x$$
with an inf\/inite number of commuting constants of
motion. The common level sets of these constants of motion are
generically inf\/inite dimensional tori of maximal dimension. These constants of the motion have
linearly independent gradients, at certain critical tori, which
become linearly dependent when the critical tori have dimension lower
than maximal (but otherwise arbitrary). Critical tori can be either stable or
unstable. 
In the unstable case, the constant level sets have a saddle structure in a
neighborhood of the critical torus. An unstable critical torus has an unstable
manifold of inf\/inite dimension. If, a phase point on this manifold
approaches the critical torus as $t\lgh \pm\infty$, such a phase point is said
to lie on an orbit which is homoclinic to the critical torus.

Bishop {\it et al.} \cite{bf1} have studied the sine--Gordon 
attractor numerically for $g_{2}(t)={\Gamma}{\cos}{\o}t$ and 
the parameter region $0.1\le {\e}{\Gamma}, {\e}a \le
1$. In that region the so--called ``breather'' solutions (i.e localised in
space and periodic in time) interact and
qualitatively the dynamics appears to be well described by a four mode
truncation \cite{halwig}, \cite{kw}. 
 
Ercolani {\it et al.,} \cite{erc1} have explored the existence of homoclinic
orbits 
for the unperturbed system. To this end, they used the fact that
system (\ref{1.2}) admits a 
Lax pair representation and derived explicit formulas for all homoclinic
orbits 
in terms of B\"acklund transformations. In these formulas, the use of the 
imaginary double points of the spectrum of Lax operator is very important,
since 
it describes the homoclinic orbits in inf\/inite--dimensional space.

However, it is very important to study the behavior of the trajectories of
the perturbed problem (\ref{1.2a}), (\ref{1.2b}). Especially, it is interesting to ask:
 under which perturbations do the homoclinic 
orbits persist and how does one establish their persistence? In the
f\/inite--dimensional system, the method which is most often used to answer such questions 
is the one established by Mel'nikov \cite{mel}, \cite{wig1}.

In this paper, we extend and apply this geometric method for the
perturbed system
(\ref{1.2a}), (\ref{1.2b}), to establish the persistence of homoclinic
orbits through the transversal intersection of invariant manifolds
of singular f\/ixed point. 
In section 2, we describe 
the symplectic structure and brief\/ly outline
the theory of the $\e=0$ (integrable) case.
In fact, in the integrable setting, the associated spectral problem has been
used to classify all instabilities and to construct representations of their
homoclinic orbits and we derive an analytic expression for the
gradient of the Floquet discriminant.
Section 3 is devoted to describe the invariant manifolds for the
perturbed system (\ref{1.2a}), (\ref{1.2b}). We study the spectrum of
the linearised system in the neighborhood of the uniform solution
$u=(0, 0)$ lies on the circle $S$. The linearized operator admits two
pair of real eigenvalues ${\l}^{\pm}_{0}, {\l}^{\pm}_{1}$ and inf\/inite
number of imaginary eigenvalues ${\l}^{\pm}_{j}={\pm}{\iu}a_{j}$,
$j{\geq} 2$, thus the uniform solution $(0, 0)$ of the sine-Gordon
equation in the whole space becomes a singular point of type
saddle-focus. We rewrite our system in a more convenient form and prove an invariant
manifold theorem for the perturbed sine-Gordon equation
(\ref{1.2a}). Also, we construct the corresponding Poincar\'e map for
the time-periodic perturbed system (\ref{1.2b}) and analyze the dynamics on the
perturbed manifolds.
%as a system of integral equations and introduce a sup norm in the functional
%space to verify that our system admits a unique solution. 

We derive, in section 4, a formula for the distance between
the codimension two center-stable and unstable manifold of the
critical torus $S$ of the 
perturbed system. We prove the existence of persistent homoclinic
orbit to the saddle-focus point $O_{\e}=O+{\Or}(\e)\in S$ for the system
(\ref{1.2a}) and for (\ref{1.2b}) with $g_{2}(t)={\G}{\cos}{\o}t$, this problem has been
studied numerically by Bishop {\it et al.} \cite{bf1}. 
%The
%perturbation $g(t, u)$ can be quadratic in $u$, however this
%assumption requires a normal form transformation to shift the
%nonlinearity in higher order terms. This idea will be fully developed
%in the reference, \cite{r14}.  
The derivation of the distance between the manifolds 
is an extension of Mel'nikov methods in inf\/inite 
dimensional space. We prove the convergence of the Mel'nikov integrals
in Case I and II. The existence of simple zeros of the
distance function implies a transversal intersection of those
manifolds, as long as certain additional non--degeneracy conditions 
are satisf\/ied. These simple zeros correspond to orbits that are
asymptotic to the saddle-focus point ${O}_{\e}$ in backward time
following the curve which lies in the unstable manifold and in forward
time following the curves on the center-stable manifold.
 
\section{Discussion of the Integrable System}
\setcounter{equation}{0}

Consider the nonlinear wave equation (\ref{1.2}) on a segment of length $L$
with even periodic boundary conditions in a symplectic space 
${\w X}$ (\ref{space})
%:=\{ {\bf u}=(u, u_t): e^{{\iu}u(x+L)}=e^{{\iu}u(x)},
%\quad u_t(x+L)=u_t(x) \}$$
%where we denote as $H_{1}$ the Sobolev space of square integrable functions
%$u(x)$ whose f\/irst derivatives are square integrable
with scalar product $\langle \,.\, ,\,.\,\rangle$, endowed by a weakly
symplectic form $\O$ given by
\bq
{\O}({\bf u}, {\bf u}^{1})=\int_{0}^{L}\big (u^1_tu-u_tu^1 \big ){\id}x,\quad
{\forall} {\bf u}, {\bf u}^{1}\in {\w X}
\label{2.1}
\eq
Let us def\/ine a domain ${\w P}\subset {\w X}$ and a function $f\in C^{1}({\w
P})$ and let 
${\nabla}_{\a}f\in {\w X}$ be the gradient of $f$ with respect to the inner
product in ${\w X}$
$$Df(\a)w:=\frac{1}{L}\int_{0}^{L}{\nabla}_{\a}f(\a(x))w(x){\id}x$$
Denoting by
$$H({\bf u})=\int_{0}^{L}\left(\frac{1}{2}(u^{2}_{t}+c^2u^2_{x})+(1+{\cos}\,u)\right)
{\id}x\in
C^{1}({\w X})$$ 
the Hamiltonian vector field $V_{H}$ with Hamiltonian $H$
and the map $V_{H}:{\w P}\lgh {\w X}$ such that $V_{H}=J{\nabla}H({\bf u})$,
the equation (\ref{1.2}) takes the form 
\bq
{\di {\bf u}_{t}}=\left ( \begin{array}{cc} {\di 0} & {\di 1} \\ \\ {\di -1}
& {\di 0} \end{array} \right) 
\left ( \begin{array}{c} {\di\frac{{\d}H}{{\d}u}} \\ \\
{\di\frac{{\d}H}{{\d}u_t}} 
\end{array} \right)=\left ( \begin{array}{c} {\di v} \\ \\ {\di
c^2u_{xx}+{\sin}\, u} 
\end{array} \right)
\label{2.2}
\eq
Def\/ining the Poisson bracket 
\bq
\{\,F, G\,\}=\int_{0}^{L}\left(\frac{{\d}F}{{\d}u}\frac{{\d}G}{{\d}u_t}-
\frac{{\d}F}{{\d}u_t}\frac{{\d}G}{{\d}u}\right){\id}x
\label{2.3}
\eq
we note that the evolution of any functional $F$, under the SG flow, is
governed 
by
\bq
\frac{{\id}F}{{\id}t}=\{\,F, H\,\}
\label{2.4}
\eq
Obviously, the Hamiltonian $H$ is conserved by the SG flow. The SG equation
admits an inf\/inite family of conserved functionals in involution with respect
to the Poisson bracket (\ref{2.3}). This fact allows the SG to be solved with
the inverse scattering transform \cite{fad}.

Let us denote by $D(V_H)=\{ {\bf u}\in {\w P}, V_H({\bf u})\in {\w X} \}$
the domain of $V_H$ in $\w P$. We give the following definition which
is useful for the study of invariant manifolds, contained in section 3.
%\begin{dfn}
%A curve ${\bf u}={\bf u}(t)$, for $t\in {\w I}=[\, 0, \tau]$ is called a
%strong solution of the equation (\ref{2.2}) in the space ${\w X}$ if\/f 
%${\bf u}\in C^{1}(\w I, \w P)$, ${\bf u}(t)\in D(V_H)$ for all $t$
%and (\ref{2%.2}) is satisfied. A curve is called a weak solution 
%of (\ref{2.2}) if\/f it is the
%limit in the $C({\w I}, \w X)$--norm of some sequence of strong solutions, 
%where $C({\w I}, \w X)$ denotes the space of continuous functions $Y:{\w
%I}\rightarrow  {\w X}$.
%\end{dfn}
\begin{dfn}
Let ${\w P}_{1}\subset {\w P}$ be such that for every 
${\bf u}_0\in {\w P}_{1}$
there exists a unique solution ${\bf u}(t)={\Phi}^{t}({\bf u}_0)$, $t\in
{\w I}$ of (\ref{2.2}) with initial condition ${\bf u}_0={\bf u}(0)$. The set
of the mappings 
$${\Phi}^{t}:{\w P}_{1}\lgh {\w P},\quad {\bf u}_0\lgh {\Phi}^{t}({\bf u}_0)
\quad t\in {\w I}$$ is called the flow of equation (\ref{2.2}).
\end{dfn}

We are interested in studying the dynamics of the perturbed system (\ref{1.2})
on a submanifold, ${\tilde {\w X}}$, of codimension 2, in the
inf\/inite--dimensional  
phase space $\w X$, given by two functionally independent equations
%$${\bf u}_x(x, t)=0$$
%whereby the system (\ref{1.2}) is constrained by the relation
%${\bf u}_x(x, t)=0$. The submanifold ${\tilde {\w X}}$ def\/ined by this
%set of spatially independent functions ${\bf u}$, is written as
\bq
{\tilde {\w X}}=\Big\{\,{\bf u}\in {\w X}\,:\,{\bf u}_x(x, t)=0\,\Big\}
\label{2.5}
\eq
and (\ref{1.2}) becomes a pendulum system, for which we know its
dynamics. Thus, starting with a 2--dimensional subspace $\Pi$ of the 
codimension 2
manifold ${\tilde {\w X}}$, we are in a position to follow the trajectories of
(\ref{1.2}) and study their behavior in the inf\/inite dimensional space.

The phase space of (\ref{1.2}) can be described in terms of the spectrum of the
linear operator (for a detailed description see \cite{erc1}, \cite{for1}):
\bq
L^{(x)}({\bf u}, \zeta):=-{\iu}c\left ( \begin{array}{cc} 0 & -{\iu} \\ {\iu} & 0
\end{array} \right)\frac{\id}{{\id}x}+\frac{\iu}{4}(cu_x+u_t)\left ( \begin{array}{cc}
0 & 1 \\ 1 & 0 \end{array} \right)-\frac{1}{16\zeta}\left (
\begin{array}{cc} e^{{\iu}u} & 0 \\ 0 & e^{-{\iu}u} \end{array}
\right)-{\zeta}I    
\label{2.6}
\eq
The SG equation arises as the compatibility conditions of the following Lax
pair of linear operators 
\bq
L^{(x)}({\bf u}, \zeta){\phi}=0,\quad
L^{(t)}({\bf u}, \zeta){\phi}=0
\label{2.7}
\eq
where 
\bq
L^{(t)}({\bf u}, \zeta):=-{\iu}\left ( \begin{array}{cc} 0 & -{\iu} \\ {\iu} & 0
\end{array} \right)\frac{{\id}}{{\id}t}+\frac{\iu}{4}(cu_x+u_t)\left ( \begin{array}{cc}
0 & 1 \\ 1 & 0 \end{array} \right)+\frac{1}{16\zeta}\left (
\begin{array}{cc} e^{{\iu}u} & 0 \\ 0 & e^{-{\iu}u} \end{array}
\right)-{\zeta}I    
\label{2.8}
\eq
$I$ is the identity matrix, ${\bf u}=(u(x, t), u_t(x, t))$ the potential and
$\zeta\in {\Cset}$ denotes the spectral parameter.

The spectrum of $L^{(x)}$, denoted by
\bq
{\sigma}(L^{(x)}):=\{{\zeta}\in {\Cset}: L^{(x)}{\phi}=0, {\vert \phi \vert}<
\infty, {\forall}x\},
\label{2.9}
\eq
characterizes the solution of the SG, the potential ${\bf u}$ also satisf\/ies
the SG and is of spatial period $L$. To achieve the spectral analysis of
$L^{(x)}$, we use Floquet Theory as follows \cite{for1}:

Let us consider the fundamental matrix $M(x, x_0; \bf u, \zeta)$ of $L^{(x)}$
as:
\bq
L^{(x)}({\bf u}, \zeta)M=0, \quad M(x_0, x_0; {\bf u}, \zeta)=I
\label{2.10}
\eq
and the Floquet discriminant 
$${\D}({\bf u}, \zeta):={\trace}M(x_0+L, x_0; {\bf u}, \zeta)$$
The spectrum of $L^{(x)}$ is given by the following condition
\bq
{\sigma}(L^{(x)}):=\Big\{\,{\zeta}\in {\Cset}\,:\,{\D}({\bf u},
\zeta)\in {\Rset},\quad {\vert {\D}({\bf u}, \zeta) \vert}\leq 2\,\Big\}
\label{2.11}
\eq
where the discriminant ${\D}$ is known to be analytic in both its
arguments and is invariant along solutions of the SG equation:
$$\frac{{\id}}{{\id}t}{\D}({\bf u}(t), \zeta)=0$$
Thus for each $\zeta$ ${\D}$ is a sine-Gordon constant 
of the motion. Moreover, by varying $\zeta$, we generate all 
of the sine-Gordon constants. Thus, the level sets 
${\rm M}_{\bf u}$ are def\/ined through $\D$:
\[
{\rm M}_{\bf u}=\Big\{\, {\bf v}\in {\w X}\,:\, 
{\D}({\bf u}, \zeta)={\D}({\bf v}, \zeta),\quad 
{\forall}{\zeta}\in {\Cset}\, \Big\}
\]
We have the following elements of the ${\sigma}(L^{(x)})$ which determine the
nonlinear mode content of solutions of sine--Gordon equations and the dynamical
stability of these modes:
\begin{itemize}
\item[(i)] simple periodic (antiperiodic) spectrum ${\sigma}^{s}_{+} ({\sigma}^{s}_{-})$
$${\sigma}^{s}_{\pm}=\Big\{{\zeta}^{s}\in {\Cset}\,:\,{\D}({\bf u},
\zeta)\Big|_
{{\zeta}^{s}={\zeta}^{s}_{\pm}}=\pm 2, \quad
\frac{{\id}{\D}}{{\id}{\zeta}}\Big|_
{{\zeta}^{s}={\zeta}^{s}_{\pm}}\neq 0\,\Big\}$$
\item[(ii)] double points of the simple periodic (antiperiodic) spectrum
$${\sigma}^d=\Big\{{\zeta}^{d}\in {\Cset}\,:\,{\D}({\bf u}, \zeta)
\Big|_{{\zeta}^{d}}=\pm 2,\quad 
\frac{{\id}{\D}}{{\id}{\zeta}}\Big|_{{\zeta}^{d}}=0,\quad 
\frac{{\id}^2{\D}}{{\id}{\zeta}}\Big|_{{\zeta}^{d}}\neq 0\,\Big\}$$
\end{itemize}
One studies the stability of a spatially indepedent 
solution $u(x, t)={\rm c}(t)$ of the sine-Gordon 
equation, a solution which is periodic in $t$. 
%All the pendula swing together, with no spatial depedence. 
As is well known \cite{gh}, an analytical formula for this solution is 
\bq
u(x, t)={\rm c}(t)=2{\sin}^{-1}\big(m{\sn}(t; m)\big)
\label{new1}
\eq
where ${\sn}(t; m)$ is the Jacobi elliptic function, and 
its modulus $m$ measures the amplitude of oscillation. 
Since ${\rm c}(t)$ is indepedent of $x$, the 
linearization can be solved by a 
Fourier transformation from $x\lgh k=k_{j}=2{\pi}j/L$,
\bq
{\w U}(x, t)=\sum_{j\in {\Zset}}{\exp}[{\iu}k_{j}x]\,
{\hat {\w U}}(k_{j}, t)
\label{new2}
\eq
This analysis is yields the following result: modes 
${\hat {\w U}}(k_{j}, t)$ are exponentially 
growing in $t$ if\/f $k^{2}_{j}\in (\,0, m^{2})$. 
That is, the $x$-indepedent solution 
${\rm c}(t)$ is unstable to long-wavelength 
perturbations. 
The number $K$ of unstable modes increases 
with the period $L$. We list the unstable modes
\[
2{\pi}/L, 2\big(2{\pi}/L\big),\ldots, 
K\big(2{\pi}/L\big)=\big[ m \big]
\]
where $\big[ m \big]$ denotes the largest integer 
multiple of $2{\pi}/L$.

As an example let us compute the spectrum for the spatially and 
temporally uniform
solution ${\bf u}=(0, 0)$. In this case, the f\/irst system of (\ref{2.7}) has
constant coef\/f\/icients and is readily solved. The Floquet discriminant is given
by 
\bq
{\D}({\bf u}, \zeta)=2{\cos}\, \Big[
\big ( {\zeta}+\frac{1}{16\zeta} \big )\frac{L}{c}\Big ]
\label{2.12}
\eq
Making use of the def\/inition above, the continuous spectrum is
given by the entire real axis as well as the curve ${\vert \zeta
\vert}^2=\frac{1}{16}$ in the complex $\zeta$--plane. The simple periodic 
spectrum is given by
\bq
{\zeta}=\frac{1}{2}\Big ( \frac{jc{\pi}}{L}\pm
{\sqrt{\big ( \frac{jc{\pi}}{L}\big )^2-\frac{1}{4}}} \Big ), \quad j\in {\Zset}
\label{2.13}
\eq
Each of these points is a double point embedded in the continuous spectrum and
becomes complex if
\bq
0\leq {\big ( \frac{2{\pi}jc}{L} \big )}^2\le 1
\label{2.14}
\eq
The condition (\ref{2.14}) is exactly the same as the condition for
linearized instability (see section 3 below) and the complex double point is given by
\bq
{\zeta}_{d}=\frac{1}{4}{\exp}[{\iu}{\b}]
\label{2.15}
\eq
with $${\b}={\tan}^{-1}\,
\frac{L}{2j{\pi}c}{\sqrt{1-\big (\frac{2jc{\pi}}{L}} \big )^2}$$
The Floquet discriminant is an invariant of the SG equation.
This means that ${\D}({\bf u}, \zeta)$ satisf\/ies the following Poisson bracket
conditions
\bq
\{{\D}({\bf u}, \zeta), {\D}({\bf u}, {\zeta}^{\prime})\}=0,\quad
{\forall} {\zeta}, {\zeta}^{\prime}\in {\Cset},\quad
\{{\D}({\bf u}, \zeta), H({\bf u})\}=0,\quad
{\forall} {\zeta}\in {\Cset}
\label{2.16}
\eq
and the grad of $\D$ is given by (see Appendix)
\bqn
\frac{{\d}{\D}}{{\d}u}({\bf u}, \zeta) & = & \frac{\iu}{4}{\trace}
\bigg [ M(L)D_{x}\big [M^{-1}(x){\w I}M(x) \big ]+\frac{1}{4c{\zeta}}M(L)M^{-1}(x)
{\w E}M(x)\bigg ] \cr\cr
\frac{{\d}{\D}}{{\d}u_t}({\bf u}, \zeta) & = & -\frac{\iu}{4c}{\trace}
\bigg [ M(L){M}^{-1}(x){\w I}
%\left ( \begin{array}{cc} 1 & 0 \\ 0 & -1 \end{array} \right) 
M(x)\bigg ]
\label{2.17}
\eqn
\[
{\w I}=\left ( \begin{array}{cc} 1 & 0 \\ 0 & -1
\end{array} \right ),\qquad {\w E}=\left (
\begin{array}{cc} 0 & e^{-{\iu}u} \\ e^{{\iu}u} & 0 
\end{array} \right )
\]
where $M(L)=M(L, x_{0}, \bf u, \zeta)$. Making use of the fact that $(0, 0)$
satisf\/ies the SG equation and the spectrum  
${\sigma}(L^{(x)})$ is invariant in $t$, provides a countable number of
constants of motion. 
Let $u(x, t)$ denote a solution of the 
SG equation which is periodic in $x$ with 
period $L$ and $U(x, t)$ denote one of the 
B\"acklund transform of $u$ as defined by 
(A.2) in the Appendix. Then their Floquet 
discriminant satisf\/ies
\[
{\D}(\zeta, U)={\D}(\zeta, u),\quad 
{\forall}{\zeta}\in {\Cset}
\]
that is $U$ and $u$ lie on the same isospectral 
set in $\w X$. When $u\in {\w X}_{NQ}$ (the $N$-phase 
quasiperiodic waves), one component of the level set 
${\rm M}_{\bf u}$ is an $N$-torus 
${\Tset}^{N}={\rm M}_{\bf u}\times{\w X}_{NQ}$. 
Let $u\in {\Tset}^{N}$ have a fully complex double 
point ${\zeta}_{d}$, as an example we return to 
the case ${\bf u}=O(=(0, 0))$ with 
${\zeta}_{d}=\frac{1}{4}e^{{\iu}{\b}}$. 
Let $U\in {\rm M}_{\bf u}$ as defined by 
\bq
U(x, t)=4{\tan}^{-1}\Big (
{\tan}\,{\b}\,{\cos}\,({x}{c}\,{\cos}\,{\b})\,{\sech}\,(t\,{\sin}\,{\b}) \Big)
\label{2.18}
\eq
Then \cite{erc1}
\begin{itemize}
\item[i.]\, $U\not\in {\Tset}^{N}$ %den anikei
\item[ii.]\, $U$ is homoclinic to ${\Tset}^{N}$, 
i.e there exists $U^{\pm}\in {\Tset}^{N}$ such that 
${\Vert U(x, t)-U^{\pm}(x, t)\Vert}\lgh 0$ as 
$t\rh {\pm}{\infty}$.
\end{itemize}
The expression (\ref{2.18}) is homoclinic to the torus, 
\bq
U(x, t)={\Or}[\,{\exp}(\,-\sqrt{1-c^2}{\vert t\vert}\,)\,],\quad 
{\rm as} \quad t\lgh \pm\infty
\label{homcon}
\eq
The solution (\ref{2.18}) is a homoclinic orbit of the integrable SG equation
and is related to the famous soliton type solution known as the
``breather'' \cite{erc1},
in the sense that it may be called a ``spatial breather'', since it is localized in $t$ and periodic in space.

\section{An Invariant Manifold Theorem}
\setcounter{equation}{0}
In this section, we state and give the proof of an invariant manifold 
theorem for the perturbed sine-Gordon equation.
There exists a two dimensional subspace ${\tilde {\w X}}$
\[
{\tilde {\w X}}=\Big\{\, u\in {\Hset}^{1}\,:\,u_{x}=0\,\Big\}
\]
such that the set ${\Pi}={\tilde {\w X}}\cap {\Hset}^{1}$ is invariant
under the f\/low of the perturbed system. The manifold $\Pi$ is real
symplectic with a nondegenerate 2-form ${\O}_{\Pi}$ which arises from
the restriction of the symplectic structure to $\Pi$. For $\e=0$,
the system (\ref{1.2}) restricted to $\Pi$ becomes a 1
degree-of-freedom completely integrable Hamiltonian system, pendulum
system, for which we know its dynamics. Furthermore, the uniform
solution 
$O=(0, 0)$ is a saddle type for the reduced pendulum system which 
lies on the torus $S$. On the whole space this uniform solution becomes
a singular point of type saddle-focus. The breather solution
$U(x, t)$ of the full system approaches this set when $t\to
{\pm}{\infty}$. Starting with the space $\Pi$, we are in a position to
follow the trajectories of (\ref{1.2a}) and study the behavior of its
solution in the inf\/inite dimensional space. 

\subsection{Linearized Analysis}

The linearized system of the perturbed sine-Gordon in the neighborhood
of the solution  $O=(0, 0)\in S$ takes the following form for the Case I;
\bq
{{\rm {\bf r}}}_{t}=A{{\rm {\bf r}}}+{\rK}_{I}({\rm {\bf r}})
\label{3.1n}
\eq
where ${\rm {\bf r}}=({\hr}, {\hr}_{t})$
$$A=\left ( \begin{array}{cc} 0 & 1 \\ c^2{\pt}^{2}_{x}+1 &
0 \end{array} \right),\quad {\rK}_{I}({\rm {\bf r}})=
\left ( \begin{array}{cc} 0 \\ -{\e}a{\hr}_{t}-
{\tilde G}_{I}({\hr}; \e) 
\end{array} \right)$$
and the function ${\tilde G}_{I}$ is def\/ined as follows:
\bq
{\tilde G}_{I}({\hr})={\e}g_{1}\,(O+{\hr})+{\sin}\,{\hr}-{\hr}=
{\Or}({\hr}^3)
\label{kfg}
\eq
for the Case II, 
\bq
{{\rm {\bf r}}}_{t}=A{\rm {\bf r}}+{\rK}_{II}({\rm {\bf r}})
\label{3.1bn}
\eq
where 
$${\rK}_{II}({\rm {\bf r}})=\left ( \begin{array}{cc} 0 \\ -{\e}a{\hr}_{t}-
{\tilde G}_{II}({\hr}; \e) 
\end{array} \right)$$
and the function ${\tilde G}_{II}$ is def\/ined as follows:
\bq
{\tilde G}_{II}={\e}g_{2}\,(t)+{\sin}\,{\hr}-{\hr}=
{\Or}({\hr}^3)
\label{kfgb}
\eq
We can also combine the two equations (\ref{3.1n}) in a single
equation, (respectively for the Case II) 
\bq
{\hr}_{tt}-c^{2}{\hr}_{xx}-{\hr}+{\e}a{\hr}_{t}=
{\tilde G}_{I}({\hr})
\label{op1}
\eq
Let $\e=0$, (\ref{op1}) becomes
\bq 
{\hr}_{tt}-c^2{\hr}_{xx}-{\hr}=0, \quad 
c^2\in \Big(\, \frac{1}{4}, 1 \Big) 
\label{3.2}
\eq
Substituting ${\hr}={\hat r}(t)e^{{\iu}k_{j}x}$ with
$k_{j}=\frac{2{\pi}j}{L}$, and $j$  
an arbitrary integer we obtain from (\ref{3.2})
$${\hat r}^{{\prime}{\prime}}-(1-c^{2}k^{2}_{j}){\hat r}=0$$
This shows that the $j$-th mode grows exponentially if
$c^{2}(2{\pi}j/L)^{2}<1, j=0,1$. Using now the even spatial periodic
boundary condition we see that 
our problem has two unstable directions, 
%for $j=0$ in the plane $\Pi$
corresponding to the unstable 
directions of the saddle point $(0, 0)$: The
zeroth mode in the plane $\Pi$ ${\alpha}_{0}=1$, which is the most unstable
in the sense that it grows at the fastest rate and $j=1$, 
${\alpha}_{1}=\sqrt{1-(2{\pi}c/L)^{2}}$ of\/f the 
subspace $\Pi$. Hence, the saddle point $O=(0, 0)$ has a
2--dimensional unstable manifold. 
If $c^{2}(2{\pi}j/L)^{2}>1$, we f\/ind inf\/inite number imaginary
eigenvalues 
${\l}_{j}={\pm}{\iu}{\alpha}_{j}={\pm}{\iu}\sqrt{\,c^{2}(2{\pi}j/L)^{2}-1\,},
j\geq 2$.

In order to study the local behavior of solutions of the nonlinear system
(\ref{3.1n}) we need to analyze the spectrum of the operator
$${\w L}_{\e}=\left ( \begin{array}{cc} 0 & 1 \\ c^2{\pt}^{2}_{x}+1 &
-{\e}a \end{array} \right)$$
We consider the eigenvalue problem ${\w L}_{\e}e={\l}e$ for the eigen--pairs 
$\{ e(x), \l \}$. Using Fourier expansions of $e(x)$ in
${\exp}({\iu}k_{j}x)$ leads to a quadratic
expression for the eigenvalues $\l$:
$${\l}^{2}+{\l}{\e}a-(1-c^{2}k^{2}_{j})=0,\quad k_{j}=2{\pi}j/L,\quad
j=0,1,2,{\ldots}$$ 
For $j=0,1$, we have:
\bqns
e_{{\st}, {\un}} & = & (\,1, \mp{\s}_{j})^{\bot}{\cos}\,k_{j}x\cr\cr
{\l}^{\e}_{{\st}, {\un}} & = & -\frac{{\e}a}{2}\pm
{\sqrt{\big ( \frac{{\e}a}{2}\big )^2+{\s}^{2}_{j}}}, \quad 
{\s}_{j}=\sqrt{1-c^{2}k^{2}_{j}}
\eqns
while for $j\geq 2$, the eigenvalues come in complex conjugate pairs.

\subsection{Invariant Manifolds-Case I}

From the above analysis, we conclude that the function ${u}$ may be
written as a linear combination 
$${u}={\hr}_{\un}{\bf e}_{\un}+{\hr}_{\st}{\bf e}_{\st}+{\hr}_{\ce}$$
and in terms of these variables the equation ${\hr}_{t}={\w L}_{e}{\hr}$ splits into 
\bqn
{\hr}_{{\un}, t} & = & {\l}^{\e}_{\un}{\hr}_{\un}\cr
{\hr}_{{\st}, t} & = & -{\l}^{\e}_{\st}{\hr}_{\st}\cr
{\hr}_{{\ce}, t} & = & {\w A}{\hr}_{\ce}
\label{3.3}
\eqn
where the operator $\w A$ corresponds to the inf\/inite number of center
directions. 
In a neighborhood of the f\/ixed point $(0, 0)$ which lies on a torus $S$,
the nonlinear equation (\ref{3.1n}) can be viewed as a perturbation
of the linear equation (\ref{3.3}). Under the f\/low of this linear equation
and for ${\e}=0$, $S$ has a 2--dimensional stable and unstable manifold together
with a codimension 4 center manifold. We focus our attention on the center
manifold $E^{c}(S)$ together with center-stable 
$E^{{\ce}{\st}}(S)$ and center-unstable $E^{{\ce}{\un}}(S)$ 
manifolds, on which the solutions have growth rates bounded by 
\begin{itemize}
\item[(i)]
$
{\exp}[\,{\s}_{j}t/n\,],\quad {\rm for} \quad t>0,\quad {\rm and} \quad 
{\exp}[\,-{\s}_{j}t/n\,],\quad {\rm for}\quad t<0, \quad j=0,1$
\item[(ii)]
${\exp}[\,{\s}_{j}{\vert t \vert}/n\,],\quad {\rm for} \quad {\forall
t},\quad j\geq 2, \quad {\s}_{j}={\iu}\sqrt{(ck_{j})^{2}-1}$
\end{itemize}
respectively, while for the operator $\w A$ we have
$${\| {\exp}[\,{\w A}t\,] \|}\leq nC{\exp}[\,{\s}_{j}{\vert t \vert}/n\,]$$
for some integer $n$ and positive constant $C$.

Let us now concentrate on the existence of locally invariant manifolds of
the perturbed SG equation. To this
end, we introduce a localization function, \cite{mar}, such that for all 
$c_1$, $c_2$ with $0< c_{1} < c_{2}$ and integer $p\geq 0$ there is a function
${\varphi}_{\d}\in C^{p}$, ${\varphi}_{\d}(y)={\varphi}(y/{\d})$:
$$
{\varphi}:{\Rset}\lgh {\Rset},\quad {\varphi}(y)=\left\{
\begin{array}{r@{\quad {\rm for}
\quad}l} 1 & {\vert y \vert}\leq c_{1} \\
0 & {\vert y \vert}\geq c_{2} \end{array} \right.
$$
with ${\sup}_{y}{\vert {\varphi}^{(\imath)}(y) \vert}\leq c$ 
for all ${\imath}\leq
p$ and suitable $c$.
Hence, equation (\ref{1.2a}) is written as a perturbation of (\ref{3.3})
with ${u}=({\hr}_{\st}, {\hr}_{\un}, {\hr}_{\ce})$
\bqn
{\hr}_{{\un}, t} & = & {\l}^{\e}_{\un}{\hr}_{\un}+P_{\un}({\hr}, {\d};\e)\cr
{\hr}_{{\st}, t} & = & -{\l}^{\e}_{\st}{\hr}_{\st}+P_{\st}({\hr}, {\d}; \e)\cr 
{\hr}_{{\ce}, t} & = & {\w A}{\hr}_{\ce}+P_{\ce}({\hr}, {\d}; \e)
\label{3.4}
\eqn
where $P_{\un}$, $P_{\st}$, $P_{\ce}$ are nonlinear functions. 
This localisation has
the ef\/fect of keeping the f\/low unchanged in the neighborhood of
torus $S$.
We are now in position to prove the existence of an inf\/inite
dimensional, locally invariant manifold for equation (\ref{1.2a}) in
the vicinity of torus $S$, local invariance means that
solutions can only leave the manifold through its boundary. The
manifold we identify for equation (\ref{1.2a}) is a codimension 4
center manifold for the torus $S$ for $\e=0$, and continuous to exist
for nonzero values, even through $S$ is usually destroyed by the
perturbations. We also show that the manifold possesses locally
invariant stable and unstable manifolds which codimension 2. 

The following theorem concerns the existence of the locally invariant
manifolds described above. 

\begin{thm}
There exists a $\d$ neighborhood of $S\subset {\Hset}^{1}$ and ${\e}_{0}(\d)>0$
such that ${\forall}\,{\e}\,\in [\,0, {\e}_{0})$ equation (\ref{3.4})
has locally invariant center-stable and unstable manifolds of codimension
2
\bqn
W^{\st}_{\rm loc}({\w M}_{\e}) & = & \big\{\, {\hr}\in
{\Hset}^{1}\,:\, 
{\hr}_{\un}\,=\,h^{\un}\,(\,{\hr}_{\st}, {\hr}_{\ce}; \e) 
\,\big\}\cr\cr 
W^{\un}_{\rm loc}({\w M}_{\e}) & = & \big\{\, {\hr}\in
{\Hset}^{1}\,:\, 
{\hr}_{\st}\,=\,h^{\st}\,(\,{\hr}_{\un}, {\hr}_{\ce}; \e)\, 
\big\}
\label{3.17}
\eqn
with ${\hr}_{\un}=({\hr}_{{\un}, 0}, {\hr}_{{\un}, 1}), 
{\hr}_{\st}=({\hr}_{{\st}, 0}, {\hr}_{{\st}, 1})$ and a codimension 4 center manifold
\bq
{\w M}_{\e}=\big\{\, {\hr}\in {\Hset}^{1}\,:\, {\hr}_{\un}\,=\,
h^{{\ce}{\un}}\,(\,{\hr}_{\ce}; \e),\, 
{\hr}_{\st}\,=\,h^{{\ce}{\st}}\,(\,{\hr}_{\ce}; \e)\,\big\} 
\label{3.18}
\eq
where ${\hr}_{\ce}$ is defined all of ${\Rset}$ and satisfies
${\sup}_{t\in {\Rset}}\,{\| {\hr}_{\ce} \|}_{{\Hset}^{1}}\leq 
\frac{{\d}^{\prime}}{2}$, 
while ${\d}^{\prime}$ belongs to a neighborhood of $\d$. 
\end{thm}
\proof
We write (\ref{3.4}) in the form of integral equations
\bqn
{\hr}_{\un}(t) & = & {\exp}[\,{\l}^{\e}_{\un}\,(t-t_{\un})\,]
{\hr}_{\un}(t_{\un})+\int_{t_{\un}}^{t} 
\,{\exp}[\,{\l}^{\e}_{\un}\,(t-\xi)\,]P_{\un}({\hr}(\xi), {\d}, \e){\id}{\xi}\cr\cr
{\hr}_{\st}(t) & = & {\exp}[\,-{\l}^{\e}_{\st}\,(t-t_{\st})\,]
{\hr}_{\st}(t_\st)+\int_{t_{\st}}^{t} 
\,{\exp}[\,-{\l}^{\e}_{\st}\,(t-\xi)\,]P_{\st}({\hr}(\xi), {\d} ,\e){\id}{\xi}\cr\cr
{\hr}_{\ce}(t) & = & {\exp}[\,{\w A}t\,]{\hr}_{\ce}(0)+\int_{0}^{t}
\,{\exp}[\,{\w A}\,(t-\xi)\,]P_{\ce}({\hr}(\xi), {\d}, \e){\id}{\xi}
\label{3.5}
\eqn
From the gap in the growth rates and the def\/inition of the f\/low ${\rF}^{t}_{\e}$,
we characterize the invariant manifolds $W^{\st}_{\rm loc}({\w M}_{\e})$, 
$W^{\un}_{\rm loc}({\w M}_{\e})$ by
\bqns
W^{\st}_{\rm loc}({\w M}_{\e}) & = & \Big\{ {\bar r}\in {\Hset}^{1}\,:\,{\sup}_{t\geq
0}\{\,{\exp}[\,-{\s}_{1}t/{\k}\,]{\| 
{\rF}^{t}_{\e}({\bar r}; \e) \|}_{{\Hset}^{1}}\}< \infty \Big\}\cr\cr
W^{\un}_{\rm loc}({\w M}_{\e}) & = & \Big\{ {\bar r}\in {\Hset}^{1}\,:\,{\sup}_{t\leq
0}\{\,{\exp}[\,{\s}_{1}t/{\k}\,]{\| 
{\rF}^{t}_{\e}({\bar r}; \e) \|}_{{\Hset}^{1}}\}< \infty \Big\}
\eqns
Focusing our attention upon $W^{{\ce}{\st}}_{\rm loc}({\w M}_{\e})$, for $\hat r$ in a sphere $B$ of
arbitrary radius $\varrho$, we introduce the norm
$$
{\| {\hr} \|}_{\mu}:=\sup_{t\geq 0}\big\{ {\exp}[\,-{\s}_{1}t/{\mu}\,]{\| {\hr}
\|}_{{\Hset}^{1}} \big\},\quad {\varrho}\leq {\| {\hr} \|}_{{\Hset}^{1}}
$$
For ${\hr}\in W^{{\ce}{\st}}_{\rm loc}({\w M}_{\e})$, we have ${\exp}[-{\l}^{\e}_{\un}t_{\un}]
{\vert {\hr}_{\un}(t_{\un}) \vert}\rightarrow 0$ as $t_{\un}\rightarrow \infty$.
Thus, on the center--stable manifold the integral equations (\ref{3.5})
can be written as: 
\bqn
{\hr}_{\un}(t) & = & \int_{\infty}^{t}
\,{\exp}[\,{\l}^{\e}_{\un}\,(t-\xi)\,]P_{\un}({\hr}(\xi), {\d}, \e){\id}{\xi}\cr
{\hr}_{\st}(t) & = & {\exp}[\,-{\l}^{\e}_{\st}t\,]
{\hr}_{\st}(t_{\st})+\int_{0}^{t} 
\,{\exp}[\,-{\l}^{\e}_{\st}\,(t-\xi)\,]P_{\st}({\hr}(\xi), {\d}, \e){\id}{\xi}\cr
{\hr}_{\ce}(t) & = & {\exp}[\,{\w A}t\,]{\hr}_{\ce}+\int_{0}^{t}
\,{\exp}[\,{\w A}\,(t-\xi)\,]P_{\ce}({\hr}(\xi), {\d}, \e){\id}{\xi}
\label{3.6}
\eqn

To prove the existence of $W^{\st}_{\rm loc}({\w M}_{\e})$, 
we use Newton's iterations. Let ${\hr}^{0}=0$ and
\bqn
{\hr}^{{\mm}+1}_{\un}(t) & = & \int_{+{\infty}}^{t}
\,{\exp}\big[\,{\l}^{\e}_{\un}(t-\xi)\big]\,
P_{\un}({\hr}^{\mm}(\xi), \d; \e){\id}{\xi} \cr\cr
{\hr}^{{\mm}+1}_{\st}(t) & = &
{\exp}\big[\,-{\l}^{\e}_{\st}t\,\big]{\hr}_{\st}+
\int_{0}^{t}
\,{\exp}\big[\,-{\l}^{\e}_{\st}(t-\xi)\big]\,
P_{\st}({\hr}^{\mm}(\xi), \d; \e){\id}{\xi} \cr\cr
{\hr}^{{\mm}+1}_{\ce}(t) & = &
{\exp}\big[\,{\w A}t\,\big]{\hr}_{\ce}+
\int_{0}^{t}
\,{\exp}\big[\,{\w A}(t-\xi)\big]\,
P_{\ce}({\hr}^{\mm}(\xi), \d; \e){\id}{\xi}
\label{3.13a}
\eqn

%For if ${\Vert {\hr}^{\mm} \Vert}_{{\k}_{0}}\leq C$ 
%\bqn
%{\Vert {\hr}^{{\mm}+1}(t)\Vert}_{{\Hset}^{1}} & {\leq} & 
%{\k}_{0}C{\exp}\big[\,{{\s}_{1}t}/{{\k}_{0}}\,\big]
%\Big(\,{\Vert {\hr}_{\st}\Vert}_{{\Hset}^{1}}+
%{\Vert {\hr}_{\ce}\Vert}_{{\Hset}^{1}}\,\Big)\cr\cr
%& + & \int_{t}^{\infty}{\exp}\big[\,
%\frac{{\s}_{1}}{2}(t-\xi)\,\big]
%{\Vert P_{\un}({\hr}^{\mm}(\xi), \d; \e)\Vert}_{{\Hset}^{1}}
%{\id}{\xi} \cr\cr
%& + & \int_{0}^{t}{\exp}\big[\,
%\frac{-{\s}_{1}}{2}(t-\xi)\,\big]
%{\Vert P_{\st}({\hr}^{\mm}(\xi), \d; \e)\Vert}_{{\Hset}^{1}}
%{\id}{\xi} \cr\cr
%& + & \int_{0}^{t}{\k}_{0}C{\exp}\big[\,
%\frac{{\s}_{1}}{2{\k}_{0}}
%(t-\xi)\,\big]{\Vert P_{\ce}({\hr}^{\mm}(\xi), 
%\d; \e)\Vert}_{{\Hset}^{1}}{\id}{\xi} 
%\label{new3}
%\eqn
We note that $P({\hr}, \d; \e)$ is a smooth function whose terms are
linear with coef\/f\/icient $\e$ or nonlinear and localised 
in a $\d$-neighborhood of $S$. If we let $P^{\prime}$ be the
derivative of $P$, we obtain the following estimate 
\[
{\Vert P({\hr}, \d; \e)\Vert}_{{\Hset}^{1}}\leq 
{\Vert P^{\prime} \Vert}\,{\Vert {\hr} \Vert}_{{\Hset}^{1}}+{\e}
\]
where ${\Vert P^{\prime} \Vert}$ is the supremum of the magnitude of 
$P^{\prime}$ which is equal to $C({\d}+{\e})$.

For if ${\Vert {\hr}^{\mm} \Vert}_{{\k}_{0}}\leq C$, we obtain from
(\ref{3.13a}) 
\bqns
{\Vert {\hr}^{{\mm}+1}(t)\Vert}_{{\Hset}^{1}} & {\leq} & 
{\k}_{0}C{\exp}\big[\,{{\s}_{1}t}/{{\k}_{0}}\,\big]
\Big(\,{\Vert {\hr}_{\st}\Vert}_{{\Hset}^{1}}+
{\Vert {\hr}_{\ce}\Vert}_{{\Hset}^{1}}+{\e}\,\Big)\cr\cr
& + & \int_{t}^{\infty}{\exp}\big[\,
\frac{{\s}_{1}}{2}(t-\xi)\,\big]
C({\d}+{\e}){\Vert {\hr}^{\mm}(\xi)\Vert}_{{\Hset}^{1}}{\id}{\xi} \cr\cr
& + & \int_{0}^{t}{\k}_{0}C({\d}+
{\e}){\exp}\big[\,\frac{{\s}_{1}}{2{\k}_{0}}
(t-\xi)\,\big]{\Vert {\hr}^{\mm}(\xi)\Vert}_{{\Hset}^{1}}{\id}{\xi} 
\eqns 
By using the bound of ${\hr}^{\mm}$ we obtain
\[
{\Vert {\hr}^{{\mm}+1}(t) \Vert}_{{\Hset}^{1}}\leq 
C\Big[\,{\Vert {\hr}_{\st}\Vert}_{{\Hset}^{1}}+
{\Vert {\hr}_{\ce}\Vert}_{{\Hset}^{1}}+
{\e}{\k}^{2}_{0}(\d+\e){\Vert {\hr}^{\mm} \Vert}_{{\k}_{0}}\,\Big]
{\exp}\big[\,{{\s}_{1}t}/{{\k}_{0}}\,\big]
\]
where the constant $C$ is independent of ${\k}_{0}, {\e}$ and $\d$. Fix
$\d=\frac{\a}{{\k}^{2}_{0}}$, where ${\a}=C/4$, we obtain for all 
${\e}<C/{4{\k}^{2}_{0}}$:
$$
{\Vert {\hr}^{{\mm}+1}(t) \Vert}_{{\k}_{0}}\leq 
C({\rho})+\frac{1}{2}{\Vert {\hr}^{\mm} \Vert}_{{\k}_{0}}
$$

Thus, we have proved that the sequence ${\hr}^{\mm}$ is well def\/ined and 
${\Vert {\hr}^{\mm} \Vert}_{{\k}_{0}}\leq 2C(\rho)$. Since, the
nonlinear term is smooth, we have a similar estimate for the
dif\/ference
\bq
{\Vert {\hr}^{{\mm}+1}-{\hr}^{\mm} \Vert}_{{\k}_{0}}
\leq \frac{1}{2}{\Vert {\hr}^{\mm}-{\hr}^{{\mm}-1} \Vert}
_{{\k}_{0}} 
\label{new4}
\eq
which implies that ${\hr}^{\mm}\rh {\hr}$ and 
\[
{\Vert {\hr} \Vert}_{{\k}_{0}}\leq 2C\big(\,{\e}+
{\Vert {\hr}_{\ce} \Vert}_{{\Hset}^{1}}+
{\Vert {\hr}_{\st} \Vert}_{{\Hset}^{1}}\,\big)
\]
We note that all terms in (\ref{3.6}) are smooth, 
which entails that $\big\{{\hr}^{\mm}\big\}$ 
is dif\/ferentiable.

We observe that the derivative $D{\hr}^{\mm}$ satisf\/ies:
\bqns
{\Vert D{\hr}^{\mm}(t) \Vert}_{{\Hset}^{1}} & < & C{\exp}\big[\,{{\s}_{1}t}/{{\k}_{0}}\,\big]+C\int_{t}^{\infty}{\exp}\Big[
\frac{{\s}_{1}}{2}(t-\xi)\Big]
{\Vert P^{\prime}D{\hr}^{\mm} \Vert}_{{\Hset}^{1}}{\id}{\xi}\cr\cr
& + & C\int_{0}^{t}{\exp}\Big[\frac{{\s}_{1}}{2{\k}_{0}}(t-\xi)\Big]
{\Vert P^{\prime}D{\hr}^{\mm} \Vert}_{{\Hset}^{1}}{\id}{\xi} 
\eqns
By using the bounds on $P^{\prime}$, we obtain
\[
{\Vert D{\hr}^{{\mm}+1} \Vert}_{{\k}_{0}}\leq 
C+\frac{1}{2}{\Vert D{\hr}^{\mm} \Vert}_{{\k}_{0}}
\]
which implies ${\Vert  D{\hr}^{\mm} \Vert}_{{\k}_{0}}
\leq 2C$.

Now, we are in position to estimate the dif\/ference 
between two terms in the sequence $\big\{ {\hr}^{\mm} \big\}$, 
${\d}{\hr}^{\mm}={\hr}^{\mm}-{\hr}^{{\mm}-1}$. We note by the mean
value theorem
\[
{\Vert P^{\prime}({\hr}^{\mm})-P^{\prime}({\hr}^{{\mm}-1}){\hr} \Vert}
_{{\Hset}^{1}}\leq C{\Vert {\d}{\hr}^{\mm} \Vert}_{{\Hset}^{1}}
{\Vert {\hr} \Vert}_{{\Hset}^{1}}
\]
We have:
\bqn
{\Vert D{\d}{\hr}^{{\mm}+1}(t) \Vert}_{{\Hset}^{1}} & {\leq} & 
C\int_{t}^{\infty}{\exp}\Big[\,\frac{{\s}_{1}}{2}(t-\xi)\,\Big]\Big\{\,
{\Vert P^{\prime}( D{\d}{\hr}^{\mm}) \Vert}_{{\Hset}^{1}}\cr\cr
& + & \big(\,{\Vert D{\hr}^{\mm} \Vert}_{{\Hset}^{1}}+
{\Vert D{\hr}^{{\mm}-1} \Vert}_{{\Hset}^{1}}
\,\big){\Vert {\d}{\hr}^{\mm} \Vert}_{{\Hset}^{1}}\,\Big\}{\id}{\xi}\cr\cr
& + &
C\int_{0}^{t}{\exp}\Big[\,\frac{{\s}_{1}}{2{\k}_{0}}(t-\xi)\,\Big]
\Big\{\,{\Vert P^{\prime}(D{\d}{\hr}^{\mm}) \Vert}_{{\Hset}^{1}}\cr\cr
& + & \big(\,
{\Vert D{\hr}^{\mm} \Vert}_{{\Hset}^{1}}+{\Vert D{\hr}^{{\mm}-1} \Vert}_{{\Hset}^{1}} 
\,\big){\Vert {\d}{\hr}^{\mm} \Vert}_{{\Hset}^{1}}\,\Big\}{\id}{\xi} 
\label{new5}
\eqn
and 
%The existence of quadratic terms in the equation implies to an
%increase in the growth
\[
{\Vert {\d}{\hr}^{\mm} \Vert}_{{\Hset}^{1}}{\Vert D{\hr}^{\mm} \Vert}
_{{\Hset}^{1}}\,{\leq}\,
{\exp}\Big[\,\frac{2{\s}_{1}}{{\k}_{0}}t\,\Big]
{\Vert {\d}{\hr}^{\mm} \Vert}_{{\k}_{0}} {\Vert D{\hr}^{\mm}
\Vert}_{{\k}_{0}}
\]
Then, we obtain
\[
{\Vert D{\hr}^{{\mm}+1}-D{\hr}^{\mm} \Vert}_{{\k}_{0}/2}\leq 
C{\Vert {\d}{\hr}^{\mm} \Vert}_{{\k}_{0}}+\frac{1}{2}{\Vert
D{\hr}^{\mm}-D{\hr}^{{\mm}-1} \Vert}_{{\k}_{0}}
\]
and sequence $\big\{ {\hr}^{\mm} \big\}$ converges in ${\rm C}^{1}$
using the ${\Vert {\cdot} \Vert}_{{\k}_{0}/2}$-norm. We repeat this 
procedure to f\/ind bounds on $\big\{ D^{j}{\hr}^{\mm}\big\}$ in the 
${\Vert {\cdot} \Vert}_{{\k}/j}$-norm with ${\k}/j>2$. 

We obtain ${\hr}\in C^{\rm r}$ for 
${\rm r}\,{\leq}\,\Big[\frac{{\k}_{0}}{2}\Big]-1$. Therefore, we
def\/ine $$
{\tilde r}_{\un}=h^{\un}({\tilde r}_{\st},
{\tilde r}_{\ce};\e)=\int_{\infty}^{0}{\exp}[\,-{\l}^{\e}_{\un}{\xi}\,]
P_{\un}({\hr}(\xi), \d; \e){\id}{\xi}$$
which is a $C^{\rm r}$ functional with 
${\Vert D{\hr}_{\un} \Vert}\leq 1/2$
and 
\[
W^{\st}_{\rm loc}({\w M}_{\e})
=\{\, {\hr}\in {\Hset}^{1}\,:\, {\hr}_{\un}\,=\,h^{\un}\,(\,{\hr}_{\st}, {\hr}_{\ce}; \e)\,\}
\]
is a $C^{\rm r}$ manifold of codimension 2, with
${\hr}_{\un}=({\hr}_{{\un}, 0}, {\hr}_{{\un}, 1})$. Exactly analogous results
are obtained for the center-unstable manifold, $W^{\un}_{\rm loc}({\w M}_{\e})$. 

The center manifold 
${\w M}_{\e}$ is described by 
$W^{\st}_{\rm loc}({\w M}_{\e})\cap W^{\un}_{\rm loc}({\w M}_{\e})$, and equivalent we need to
solve the following system 
\bq
{\hr}_{\un}=h^{\un}(\,{\hr}_{\st}, {\hr}_{\ce}; \e),\quad
{\hr}_{\st}=h^{\st}(\,{\hr}_{\un}, {\hr}_{\ce}; \e)
\label{new6}
\eq
we note that ${\Vert Dh^{{\un}, {\st}} \Vert}{\leq}1/2$ and
(\ref{new4}). From the implicit function theorem, we obtain that 
the system (\ref{new6}) has a unique $C^{\rm r}$ solution given by  
\[
{\hr}_{\un}=h^{{\ce}{\un}}_{\e}({\hr}_{\ce}; \e),\quad
{\hr}_{\st}=h^{{\ce}{\st}}_{\e}({\hr}_{\ce}; \e)
\]
\qed

\subsection{Invariant Manifolds-Case II}

The time-periodic perturbed sine-Gordon equation (\ref{1.2b}) can be
rewritten as:
\bq
{\bf u}_{t}={\rA}{\bf u}+{\rB}({\bf u})+{\e}\Big({\rC}{\bf
u}+g_{2}(t)\Big)
\label{c2.1}
\eq
with 
\[
{\rA}:=\left ( \begin{array}{cc} 0 & 1 \\ c^{2}{\pt}^{2}_{x} & 0
\end{array} \right),\quad {\rB}(u):=\left ( \begin{array}{cc} 0 \\
{\sin}u \end{array} \right),\quad {\rC}:=\left ( \begin{array}{cc} 0 &
0\\ 0 & -a\end{array} \right)
\]

We assume that $g_{2}:{\Hset}^{1}\times {\Sset}^{1}\to {\Hset}^{1}$ is
$C^{\infty}$ $T$-periodic in time where ${\Sset}^{1}={\Rset}/T$. The
associated extended autonomous system in a cylinder in the space
${\Hset}^{1}\times {\Sset}^{1}$
\bqn
{\dot u} & = & {\rA}u+{\rB}(u)+{\e}\big({\rC}u+g_{2}(\theta)\big)\cr
{\dot {\theta}} & = & 1
\label{c2.2}
\eqn
has a smooth f\/low ${\rF}^{t}_{\e}$. The geometry of the unperturbed
system (\ref{c2.1}) is the same with the case I. 

Suppose that linearized system of (\ref{c2.1}) with respect to the
singular point $p_{0}=O$
\bq
{{\rm {\bf r}}}_{t}={\w L}_{\e}{{{\rm {\bf r}}}}+{\e}g_{2}(t)
\label{c2.3}
\eq
with 
$${\w L}_{\e}:=\left ( \begin{array}{cc} 0 & 1 \\ c^2{\pt}^{2}_{x}+1 &
-{\e}a \end{array} \right)$$
has a $T$-periodic function ${\hr}(t, \e)$, such that ${\hr}(t,
\e)={\Or}(\e)$. 

For ${\e}>0$, the operator ${\rm e}^{T{\w L}_{\e}}$
has a spectrum consisting in two pair of real eigenvalues for $j=0,1$
and inf\/inite number of complex eigenvalues for $j\ge 2$, ${\rm
spec}({\rm e}^{T{\w L}_{\e}})_{j\ge 2}$, in such a way:
\bq
{\m}_{1}{\e}\leq {\rm dist}\Big({\rm spec}({\rm e}^{T{\w
L}_{\e}})_{j\ge 2}, \abs{z}=1\Big)\leq {\m}_{2}{\e}
\label{c2.4}
\eq
with ${\m}_{1}, {\m}_{2}>0$. 

We denote by ${\rP}^{\e}:{\Hset}^{1}\to {\Hset}^{1}$ the Poincar\'e
map for the f\/low ${\rF}^{t}_{\e}$ such that ${\rP}^{0}(p_{0})=p_{0}$
and $p_{\e}$ the f\/ixed points of ${\rP}^{\e}$ correspond to periodic
orbits of ${\rF}^{t}_{\e}$, near to $p_{0}=O:
p_{\e}=p_{0}+{\Or}(\e)$. 

Indeed, by (\ref{c2.3}) we have:
\bq
{\hr}(t, \e)={\rm e}^{t{\w L}_{\e}}{\hr}(0, \e)+\int_{0}^{t}{\e}{\rm
e}^{(t-\xi){\w L}_{\e}}g_{2}(\xi){\id}{\xi}
\label{c2.5}
\eq
and ${\hr}(T, \e)={\hr}(0, \e)$, ${\Vert {\hr}(t,
\e)\Vert}{\leq}{\m}_{3}{\e}$. 

We seek a solution $u(t, \e)$ of the
system (\ref{1.2b}) such that:
\bq
u(t, \e)={\rm e}^{t{\w L}_{\e}}{u}(0, \e)+\int_{0}^{t}{\e}{\rm
e}^{(t-\xi){\w L}_{\e}}g_{2}(\xi){\id}{\xi}
+\int_{0}^{t}{\e}
{\rm e}^{(t-\xi){\w L}_{\e}}\big({\rB}(u(\xi, \e))-u({\xi}, \e)\big){\id}{\xi}
\label{c2.6}
\eq
From (\ref{c2.5}) and (\ref{c2.6}) we obtain:
\bq
u(t, \e)-{\hr}(t, \e)={\rm e}^{t{\w L}_{\e}}({u}(0, \e)-{\hr}(0, \e))+
\int_{0}^{t}{\e}{\rm e}^{(t-\xi){\w L}_{\e}}\big({\rB}(u(\xi, \e))-u({\xi}, \e)\big){\id}{\xi}
\label{c2.7}
\eq
Thus, 
\[
{\Vert u(t, \e)-{\hr}(t, \e) \Vert}_{{\Hset}^{1}}{\leq}
{\m}_{4}{\e}+{\m}_{5}\int_{0}^{t}{\Vert u(\xi,\e) \Vert}_{{\Hset}^{1}}{\id}{\xi}
\]
using ${\Vert {\hr}(t, \e)\Vert}_{{\Hset}^{1}}{\leq}{\m}_{3}{\e}$. By Gronwall's
estimate we obtain ${\Vert u(t, \e) \Vert}_{{\Hset}^{1}}\leq {\m}_{6}{\e}$. 

We consider now the set $B_{\e}$ as
\[
B_{\e}=\Big\{\,{\hr}(t, \e): {\Vert {\hr}(t, \e)-{\hr}(0,\e)
\Vert}_{{\Hset}^{1}}<{\e}\Big\}
\]
and the map
\bqns
{\rP}^{\e}\,:\, B_{\e}&\to& {\Hset}^{1}\cr
u(0, \e)&\to& u(T, \e)
\eqns
We seek a fixed point of ${\rP}^{\e}$. From (\ref{c2.7}) $u(0, \e)$ is
a fixed point of ${\rP}^{\e}$ if and only if it is a fixed point of
the following map
\[
{\Gset}_{\e}\,:\,B_{\e}\to {\Hset}^{1}
\]
with 
\bq
{\Gset}_{\e}(u(0, \e))={\hr}(0, \e)+{\rL}^{-1}_{\e}\int_{0}^{T}{\rm
e}^{(T_\xi){\w L}_{\e}}\Big({\rB}(u(\xi, \e)-u(\xi, \e))\Big){\id}{\xi}
\label{c2.9}
\eq
where ${\rL}_{\e}=({\rm Id}-{\rm e}^{T{\w L}_{\e}})$ is invertible
operator and the condition 
\bq
{\Vert {\rL}^{-1}_{\e} \Vert}_{{\Hset}^{1}}\leq {\m}_{7}{\e}^{-1}
\label{c2.10}
\eq 
is inferred from (\ref{c2.4}).
The conditions (\ref{c2.9}) and (\ref{c2.10}) entail
\[
{\Vert {\Gset}_{\e}(u(0, \e))-{\hr}(0, \e)\Vert}_{{\Hset}^{1}}\leq {\e}
\]
which implies that for ${\e}\ll 1$ ${\Gset}_{\e}$ maps $B_{\e}$ to
itself. Also, ${\Gset}_{\e}$ is a contraction since, for $\e$ small
enough 
\bqns
{\Vert D{\Gset}_{\e}(u(0, \e))\Vert}_{{\Hset}^{1}}&=&{\big\Vert\,
{\rL}^{-1}_{\e}\int_{0}^{T}{\rm e}^{(T-\xi){\w L}_{\e}}\,D({\rB}(u(\xi,
\e)-u(\xi, \e))){\circ}\,D{\rF}^{\e}_{\xi}(u(0,
\e)){\id}{\xi}\,\big\Vert}_{{\Hset}^{1}}\cr\cr
&{\leq}&{\m}_{8}{\e} < 1 
\eqns
Thus, ${\Gset}_{\e}$ has a unique f\/ixed point on $B_{\e}$. 
There is an analogue result for the Poinacr\'e map
${\rP}^{\e}_{t_{0}}$ with the section ${\Hset}^{1}\times \{ t_{0}\}$
in ${\Hset}^{1}\times {\Sset}^{1}$.

In the following Lemma, we describe the geometry of the invariant
manifolds for the non-autonomous perturbed sine-Gordon equation
(\ref{1.2b}). Its proof based on the invariant manifolds theory
\cite{hps} and the smoothness of the flow operator ${\rF}^{t}_{\e}$. 
%\vskip 0.4 true cm
%\begin{figure}[t]
%\vspace{2cm}
%%\special{psfile=f2.ps hoffset=70 voffset=-145 vscale=60 hscale=60}
%\par
%\vspace{0.3cm}
%\begin{center}
%Figure 4. The geometry of the sine-Gordon system with quasiperiodic
%perturbation. 
%\end{center}
%\end{figure}
%\par\noindent 
%\vskip 0.4 true cm
\begin{lem}~\label{lem:lemII}
Let $p_{\e}(t_{0})$ denote the unique f\/ixed point of
${\rP}^{\e}_{t_{0}}$. Corresponding to the spectrum of ${\rm e}^{t{\w
L}_{\e}}$, there are unique invariant manifolds
$W^{{\ce}{\st}}(p_{\e}(t_{0}))$ (center-stable) and
$W^{\un}(p_{\e}(t_{0}))$ (unstable manifold) for the map
${\rP}^{\e}_{t_{0}}$ such that they are invariant under the
${\rP}^{\e}_{t_{0}}$, tangent to the eigenspaces of ${\rm e}^{t{\w
L}_{\e}}$ respectively at $p_{\e}(t_{0})$ and $C^{r}$-close as
${\e}\to 0$ to the homoclinic orbit $U(x, t)$ of the sine-Gordon
equation (\ref{1.2}) for $t_{\st}, t_{\un}$ such that
$t_{\st}<t<{\infty}, -{\infty}<t<t_{\un}$, respectively. 
We denote by $q_{\e}(t)=(p_{\e}(t), t), t\in (0, T)$ the periodic
orbit of the extended system (\ref{c2.2}) with $q_{\e}(0)=(p_{\e},
0)$. The invariant manifolds for the periodic orbit $q_{\e}(t)$ are
denoted $W^{{\ce}{\st}}(q_{\e}(t)), W^{\un}(q_{\e}(t))$ and 
\[
W^{j}(p_{\e}(t_{0}))=W^{j}(q_{\e}(t))\cap \big( {\Hset}^{1}\times \{
t_{0} \}\big),\quad {j={\ce}{\st}, {\un}}
\]
\end{lem}

\section{Mel'nikov Analysis}
\setcounter{equation}{0}
In this section, we shall combine the invariant manifold theory of the
previous section with
explicit global information from the unperturbed sine--Gordon equation to
establish the existence of persistent homoclinic orbit to the singular
point $O_{\e}$. 

\subsection{Geometry of Mel'nikov Method-Case I}

We recall that the unstable manifold $W^{\un}(O_{\e})\subset
W^{\un}({\w M}_{\e})$
is two-dimensional with one direction in the plane of 
spatial-independent solutions $\Pi$ and the other direction off the
plane in the function space ${\Hset}^{1}$. We consider an orbit in
$W^{\un}(O_{\e})$ that leaves $O_{\e}$ and takes of\/f away from the
plane, then def\/ines an orbit which lies on the codimension 2
center-unstable manifold. We will construct a distance function 
${\w D}$ between the stable and unstable manifolds, whose zeros
correspond to orbits that do not lie on the plane $\Pi$ and are
asymptotic to the saddle-focus point $O_{\e}$ in backward and forward time. 
%It is interesting to note that even through our analysis concerns a
%PDE, the geometry of the problem and the concepts introduced above
%permit us to develop a Mel'nikov theory that proceeds along the same
%lines as the corresponding one for ODEs.

Let us rewrite equation (\ref{1.2a}) in the vector form
\bq
{\bf u}_{t}\,=\,J\,{\nabla}_{{\bf u}}H({\bf u})\,+\,{\e}{\rG}_{I}
(\,{\bf u}, \mu\,)\label{4.1}
\eq
where ${\rG}_{I}(\,{\bf u},\mu\,)\,=\,( 0,
bg_{1}(u)-au_{t}\,))^{\bot}$ and ${\mu}=(a, b, c)$

For ${\e}\leq {\e}_{0}$, (\ref{3.4}) is related to the original system
(\ref{1.2a}) within some f\/ixed open set 
\[
Y=\Big\{\,({\hr}_{\st}, {\hr}_{\un}, {\hr}_{\ce})\in {\Hset}^{1}\,:\, 
{\vert {\hr}_{\st} \vert}< C_{\st},\, {\vert {\hr}_{\un} \vert}< C_{\un},\, 
{\Vert {\hr}_{\ce} \Vert}_{{\Hset}^{1}}< C_{\ce}
\Big\}
\]
where $C_{\st}, C_{\un}, C_{\ce}$ are fixed positive constants. 
We are interested in the dynamics of the solution 
${u}_{\e}(t)=( {\hr}_{\st}, {\hr}_{\un}, {\hr}_{\ce})$ of
(\ref{3.4}) in forward and backward time. We def\/ine a 
neighborhood ${\w U}$ of ${\w M}_{\e}$
\[
{\w U}=\{\,({\hr}_{\st}, {\hr}_{\un}, {\hr}_{\ce})\in Y: 
{\vert {\hr}_{\st} \vert}<{\delta},\, {\vert {\hr}_{\un}
\vert}<{\delta},\,
{\Vert {\hr}_{\ce} \Vert}_{{\Hset}^{1}} < {\delta}\leq 
C_{{\hr}_{\ce}} \}
\]
and its boundary ${\pt}{\w U}$ consists of two parts 
${\pt}{\w U}_{\st}$ stable and ${\pt}{\w U}_{\un}$ unstable as follows
\bqns
{\pt}{\w U}_{\st} & = & \big\{\,({\hr}_{\st}, {\hr}_{\un}, {\hr}_{\ce})\in {\w U}\,:\, 
{\vert {\hr}_{\st} \vert}={\delta}\big\} \cr\cr
{\pt}{\w U}_{\un} & = & \big\{\,({\hr}_{\st}, {\hr}_{\un}, {\hr}_{\ce})\in {\w U}\,:\, 
{\vert {\hr}_{\un} \vert}={\delta}\,\big\} 
\eqns
where the solution curves enter through ${\pt}{\w U}_{\st}$ and they
exit through ${\pt}{\w U}_{\un}$. 
As was said before, the breather solution $U(x, t)$ is also
a solution of the SG and is homoclinic to the point $O=(0, 0)\in
{\Pi}$ which has two unstable directions, (cf. section 3).  
%We recall that
%$\Pi$ is a two dimensional subset of the set of 
%spatially independent functions ${u}$. 
The unperturbed system possesses a
homoclinic orbit ${u}$ which evolves along the unstable direction from
the 
plane $\Pi$, intersects the unstable boundary ${\pt}{\w U}_{\un}$ at the point
${u}^{\un}_{0}$ and defines a unique orbit $(\,{u}^{\un}(t), t)$ which
lies on 
the codimension two center-unstable manifold. After a f\/inite time, this homoclinic
orbit ${u}$ intersects the stable boundary ${\pt}{\w U}_{\st}$ at the point
${u}^{\st}_{0}$ and comes back to the plane $\Pi\subset S$ and is
asymptotic to the singular point $O$, 
following a unique curve 
$({u}^{\st}(t), t)$ which lies on the center-stable manifold. Thus, for ${\e}\neq
0$, there exists a perturbed orbit ${u}^{u}_{\e}$ which evolves from the
perturbed saddle point $O_{\e}=O+{\Or}(\e)\in {\Pi}$ and intersects the unstable
boundary at the point ${u}^{\un}_{\e}$. 

Now, choose a point ${u}_{h}$ on the homoclinic orbit for the unperturbed
system. Let $\S$ be a codimension 2 hyperplane at ${u}_{h}$ which is
transversal to the homoclinic orbit $u$ and which contains the vector 
${\vec {\nu}}=(\frac{{\d}{\D}}{\d u}, \frac{{\d}{\D}}{\d u_{t}})$.
%since the discriminant
%constant of motion $\D$ (cf. (\ref{2.12}), 
%Poisson commutes with the integrable
%Hamiltonian), def\/ining the homoclinic orbit.
$W^{{\st}}({\w M}_{0})\cap {\S}$ has a codimension 1 in $\S$ and ${\vec {\nu}}$ is
transversal to $W^{{\st}}({\w M}_{0})\cap {\S}$.
Let $u_{\st}$ be the intersection of the line through 
$u_{\un}\in W^{\un}(O_{\e})\subset W^{\un}({\w M}_{\e})$
and the manifolds $W^{{\st}}({\w M}_{\e})$ 
along to direction ${\vec {\nu}}$. We may thus def\/ine the inner product
\bq
{\w D}=\langle {\nabla}{\D}, u_{\un}-u_{\st} \rangle,
\label{4.2}
\eq
as a measure of the distance between $u_{\un}$ and $u_{\st}$.

To calculate ${\w D}$, we parametrise the orbits as follows,
\[
{\rm for}\quad t{\leq} 0,\quad u=u_{h}(t+t_{*}),\quad 
u^{\un}=u^{\un}_{\e}(t+t_{*})\]
\[
{\rm for}\quad t{\geq} 0,\quad u=u_{h}(t+t_{*}),\quad 
u^{\st}=u^{\st}_{\e}(t+t_{*})
\] 
where $u^{\st}, u^{\un}$ be the solutions of the localized equations 
(\ref{3.4}). Since $u_{h}$ remains in a $\d$-neighborhood of $S$ for
$t\geq t_{*}$ $u$ is a solution of the localized system (\ref{3.4}),
we recall that the invariant manifold theorem holds in a
$\d$-neighborhood of $S\subset {\Hset}^{1}$. For $t{\leq} -t_{*}$
both orbits $u, u^{\un}$ remain in a $\d$-neighborhood of $S$. We
apply the Gronwall's estimate to system (\ref{3.4}) and from 
${\Vert u^{\un}-u_{h} \Vert}_{{\Hset}^{1}}\leq C_{1}{\e}$, for $t<0$ we have, 
\bq
{\Vert u^{\un}-u_{h} \Vert}_{{\Hset}^{1}}\leq {\e}C_{2}{\rm
 e}^{-{\s}t}
\label{u1}
\eq
Similar, for $t\geq 0$ 
\bq
{\Vert u^{\st}-u_{h} \Vert}_{{\Hset}^{1}}\leq {\e}C_{2}{\rm
 e}^{{\s}t}
\label{u2}
\eq 
The above analysis allows to decompose (\ref{4.2}) as follows;
$${\w D}=\langle {\nabla}{\D}(u_{h}), u_{\un}-{u}_h \rangle-
\langle {\nabla}{\D}(u_{h}), u_{\st}-{u}_h \rangle.$$

Let 
$$
{\w D}^{\st}=\langle {\nabla}{\D}, {u}^{\st}_{\e}(t)-{u}(t)
\rangle,\quad {t\geq 0}$$
$${\w D}^{\un}=\langle {\nabla}{\D}, {u}^{\un}_{\e}(t)-{u}(t)
\rangle,\quad {t\leq 0}$$
Then 
$${\w D}={\w D}^{\un}(0)-{\w D}^{\st}(0)$$

\begin{prop}
The distance function (\ref{4.2}) is given by 
\bq
{\w D}={\w D}^{\un}(0)-{\w D}^{\st}(0)={\e}\,M_{I}+{\Or}({\e}^2)
\label{4.3}
\eq
where 
\bq
M_{I}=\int_{-\infty}^{\infty}\langle {\nabla}{\D}(u(t)), {\rG}_{I}(u(t))
\rangle{\id}t,
\label{4.4}
\eq
denote the Mel'nikov integral with $\langle a, b
\rangle=\int_{0}^{L}a(x) b(x){\id}x$.
\end{prop}

{\proof} We start the proof by asserting from (\ref{4.1}) that 
${u}={u}(t)$, ${u}^{\un}_{\e}={u}^{\un}_{\e}(t)$,  
${u}^{\st}_{\e}={u}^{\st}_{\e}(t)$ solve the following system of equations
\bqn
{\pt}_{t}{u} & = & J\,{\nabla}H({u})\cr
{\pt}_{t}{u}^{\un}_{\e} & = & J\,{\nabla}H({u}^{\un}_{\e})+{\e}{\rG}_{I}({u}^{\un}_{\e}, \mu)\cr
{\pt}_{t}{u}^{\st}_{\e} & = &
J\,{\nabla}H({u}^{\st}_{\e})+{\e}{\rG}_{I}
({u}^{\st}_{\e}, \mu)
\label{4.5}
\eqn
Dif\/ferentiating ${\w D}^{\un}(t)$, we
have 
\bqns
{\dot {\w D}}^{\un}_{i}(t) & = & 
{\e}\langle \,{\nabla}{\D}({u}),\, {\rG}_{I}({u})\,
\rangle+\langle \,{\nabla}^2{\D}({u})\,J\,{\nabla}H({u}),\, 
{u}^{\un}_{\e}-{u}\, \rangle \cr 
& + & \langle \,{\nabla}{\D}({u}),\, 
{\nabla}(J\,{\nabla}H({u}))\,( {u}^{\un}_{\e}-{u})\,\rangle
+\langle {\nabla}{\D}(u),\, {\w S}(u^{\un}_{\e}, u)\,\rangle
\eqns
where 
\[
J\,{\nabla}H({u}^{\un}_{\e})=J\,{\nabla}H({u})+
{\nabla}(J\,{\nabla}H({u}))\,( {u}^{\un}_{\e}-{u})+{\w
S}(u^{\un}_{\e}, u)
\]
we leave out the argument $t$
of the functions for the sake of a less cumbersome notation.
From (\ref{2.15}) we have $\langle J{\nabla}H,
{\nabla}{\D}\rangle=0$, dif\/ferentiating with respect to $u$ gives
\bq
{\nabla}(J{\nabla}H)^{\bot}\,{\nabla}{\D}+{\nabla}({\nabla}{\D})\,
J{\nabla}H=0
\label{exp1}
\eq
where $^{\bot}$ denotes the matrix transpose. Taking
the inner product of (\ref{exp1}) with ${u}^{\un}_{\e}-{u}$ gives
$$\langle {\nabla}^2{\D}J\,{\nabla}H, {u}^{\un}_{\e}-{u} \rangle + 
\langle {\nabla}{\D},\, 
{\nabla}(J\,{\nabla}H)\,(\,{u}^{\un}_{\e}-{u}\,)\,\rangle=0$$ 
and thus, 
\bq
{\dot {\w D}}^{\un}(t)={\e}\langle \,{\nabla}{\D}({u}),\, 
{\rG}_{I}({u})\, \rangle + \langle \,{\nabla}{\D}({u}),\,
{\w S}(u^{\un}_{\e}, u)\,\rangle
\label{4.7}
\eq
There exists $t_{*}>0$ for $t\in (-\infty, t_{*})$ such that 
\bq
{\Vert {\nabla}{\D}({u}(t)) \Vert}_{{\Hset}^{1}}< C{\rm
e}^{{\s}t},\quad {\s}:={\sqrt{1-(2{\pi}c/L)^{2}}}
\label{u3}
\eq
since the ${\nabla}{\D}$ is evaluated on the homoclinic solution which
satisfies the condition (\ref{homcon})
and (\ref{u1}) entails 
\bq
{\Vert {\w S} \Vert}_{{\Hset}^{1}}\leq {\Vert
%c^{2}(x^{s}_{\e}-x)+{\sin}x^{\st}_{\e}-{\sin}x-(x^{\st}_{\e}-x){\cos}x+{\e}
%a(x^{\st}_{{\e},
%t}-x_{t}) \Vert}_{{\Hset}^{1}}\sim
u^{\un}-u \Vert}^{2}_{{\Hset}^{1}}\leq C_{1}{\rm e}^{-2{\d}t}{\e}^2
\label{u4}
\eq
with $2{\d}<{\s}$. Now, we want to f\/ind out the limit ${\lim}_{t\to
-{\infty}}{\w D}^{\un}(t)$ in order to represent ${\w D}(0)$ by 
\bq
{\w D}^{\un}(-\infty)+\int_{-\infty}^{0}{\dot {\w D}^{\un}}(t){\id}t
\label{u5}
\eq
We calculate ${\lim}_{t\to -\infty}{\w D}^{\un}(t)$ and show that 
$\lim_{t\to -\infty}{\w D}^{\un}(t)=0$. We have 
\[
\lim_{t\to -\infty}{\w D}^{\un}(t)=\lim_{t\to -\infty}\langle 
{\nabla}{\D}(u(t)),\, u^{\un}_{\e}(t)-u(t)\,\rangle
\]
There are $C_{2}$, $C_{3}>0$ such that 
\bqn
{\Vert u^{\un}_{\e}({\tau}-t^{\un}_{\e})-{\rF}^{\tau}_{\e}(u)\Vert}
_{{\Hset}^{1}}&{\leq}& C_{2}{\rm e}^{{\l}_{1}{\tau}}\cr\cr
{\Vert u({\tau}-t^{\un}_{0})-{\rF}^{\tau}_{\e}(u)\Vert}
_{{\Hset}^{1}}&{\leq}& C_{3}{\rm e}^{{\l}_{1}{\tau}}
\label{u6}
\eqn
for all ${\tau}\in (-\infty, 0]$, where ${\l}_{1}>0$, 
$u^{\un}_{\e}(-t^{\un}_{\e})=u^{\un}_{\e}$, $u(-t^{\un}_{0})=u^{\un}$
and 
\bq
{\rF}^{\tau}_{\e}(u)\to O_{\e}\quad {\rm as}\quad {\tau}\to -\infty
\label{u7}
\eq
\bq
{\rF}^{\tau}_{0}(u)=O\in S
\label{u8}
\eq
By these relations (\ref{u6}-\ref{u8}), we obtain that as $t\to
-\infty$
\bq
{\Vert u^{\un}_{\e}(t)-u(t)\Vert}_{{\Hset}^{1}}\sim {\Or}(t)
\label{u9}
\eq
From relations (\ref{u3}) and (\ref{u9}) we see that 
\[
\lim_{t\to -\infty}{\w D}^{\un}(t)=0
\]

All this f\/inally yields upon integration of (\ref{4.7})
\bq
{\w D}^{\un}(0)={\e}\int_{-\infty}^{0}\langle
{\nabla}{\D}({u}(t)),\, 
{\rG}_{I}({u}(t)) \rangle{\id}t+{\Or}({\e}^2)
\label{4.7a}
\eq
Similarly, we have
\bq
{\w D}^{\st}(0)=-{\e}\int_{0}^{\infty}\langle
{\nabla}{\D}({u}(t)),\, 
{\rG}_{I}({u}(t)) \rangle{\id}t+{\Or}({\e}^2)
\label{4.8}
\eq
whence the distance function (\ref{4.3}) is given by
\bq
{\w D}={\w D}^{\un}(0)-{\w D}^{\st}(0)={\e}\,M_{I}+{\Or}({\e}^2)
\label{4.9}
\eq
with 
\bq
M_{I}=\int_{-\infty}^{\infty}\langle {\nabla}{\D}(u(t)), 
{\rG}_{I}(u(t)) \rangle{\id}t,\quad {i=1,2} 
\label{4.10}
\eq
where $u(t)=U(x, t)$ the homoclinic solution of the unperturbed
system. 

The Mel'nikov integral is convergent. Because, the
perturbation term ${\rG}_{I}=bg_{1}(u)-au_{t}$ is exponential
decreasing expression, since we have assumed that $g_{1}(u)$ satisf\/ies
the conditions $g_{1}(0)=0, {\pt}_{u}g_{1}(0)=0$ and in the
neighborhood of zero 
\bq
{\Vert g_{1}(u)\Vert}_{{\Hset}^{1}}{\leq}{\mu}_{9}{\Vert
u\Vert}^{2}{\leq}{\mu}_{9}{\rm e}^{-2{\s}t}
\label{g1}
\eq
evaluated on the homoclinic orbit, with ${\mu}_{9}>0$. Also, hold
\[
{\Vert u_{t} \Vert}_{{\Hset}^{1}}{\leq}{\mu}_{10}{\rm e}^{-{\s}t}
\]
and (\ref{u3}). Thus, obtaining convergent integral $M$.
\qed
%We recall that there is a 2--dimensional unstable manifold and a
%codimension 1 center stable manifold for the system (\ref{1.2}) with
%even periodic boundary conditions. 
Using the integral of motion, we may derive for the
case I, explicit formula for the Mel'nikov function
\bq
M_{I}(\,{\b}, {\mu}\,)=\int_{-\infty}^{\infty} \langle
\,{\nabla}{\D}(U(x, t)), {\rG}_{I}(U(x, t))\, \rangle{\id}t
\label{4.11}
\eq
evaluated on the homoclinic orbit $U(x, t)$.
Thus, 
\bq
M_{I}(\,{\b}, {\mu}\,)=
b\int_{-\infty}^{\infty}\int_{0}^{L}\frac{{\d}{\D}}{{\d}u_{t}}(U){g}_{1}(U){\id}x{\id}t
-a\int_{-\infty}^{\infty}\int_{0}^{L}\frac{{\d}{\D}}{{\d}u_{t}}(U)U_{t}{\id}x{\id}t
\label{4.12}
\eq
with $\b$ defined in (\ref{2.15}) and 
\bq
\frac{{\d}{\D}}{{\d}u_{t}}=
-\frac{\iu}{4c}\bigg(\frac{1}{2{\sin}2{\b}}\bigg)^{2}
\Bigg[{\rE}\frac{{\psi}_{2}}{{\psi}_{1}}
+{\rH}\frac{{\psi}_{1}}{{\psi}_{2}}\,\Bigg]
\label{B.a}
\eq
where ${\rE}, {\rH}$ are smooth functions of the general solutions
${\psi}_{1}, {\psi}_{2}$ of the Lax pair (\ref{2.7}).

Setting $M_{I}(\,{\b}, {\mu}\,)=0$, we obtain an algebraic equation for parameters
\bq
b-a{\kappa}=0
\label{4.11s}
\eq
where 
\[
{\kappa}(\b)=\int_{-\infty}^{\infty}\int_{0}^{L}\frac{{\d}{\D}}{{\d}u_{t}}(U)U_{t}{\id}x{\id}t\,
\Bigg(\,\int_{-\infty}^{\infty}\int_{0}^{L}\frac{{\d}{\D}}{{\d}u_{t}}(U){g}_{1}(U){\id}x{\id}t\,\Bigg)^{-1}
\]
We denote the surface def\/ined by (\ref{4.11s}) by ${\w E}_{\b}$
\[
{\w E}_{\b}\,:\, {\b}_{0}={\w B}(b_{0}, a_{0}; c)
\]
There exists a region ${\w R}$ of the surface ${\w E}_{\b}$ such that 
\bq
\frac{{\pt}M_{I}}{{\pt}{\b}_{0}}({\b}_{0}, b_{0}, a_{0}; c)\neq 0
\label{4.11m}
\eq
and ${\Big \| \frac{{\pt}M_{I}}{{\pt}{\b}_{0}}\Big \|}< l,$ for
$l>0$. 
By the implicit function theorem there is a
neighborhood ${\hat {\w R}}$ of $(b_{0}, a_{0}; c)$ and a unique
$C^{n-2}$ function 
\[
{\b}_{0}={\w B}(b_{0}, a_{0}; {\e}, c)
\]
defined in ${\hat {\w R}}$ such that 
\[
{\hat {\w B}}(b_{0}, a_{0}; 0, c)={\hat {\b}}_{0}
\]
and ${\w D}({\hat {\w B}}(b_{0}, a_{0}; {\e}, c), b_{0}, a_{0}; {\e},
c)=0$. Since ${\w D}$ (the distance of stable and unstable manifolds)
is $C^{n-2}$ smooth function by (\ref{4.11m}), we obtain
\[
\frac{\pt}{{\pt}{\b}_{0}}{\w D}({\hat {\w B}}(b_{0}, a_{0}; {\e}, c), b_{0}, a_{0}; {\e},
c)\neq 0
\]
and ${\Big \| \frac{{\pt}{\w D}}{{\pt}{\b}_{0}} \Big \|}< m, m>0$, for
$(b_{0}, a_{0}; {\e}, c)\in {\hat {\w R}}$. 
Then, $W^{\st}({\w M}_{\e})$ and $W^{\un}({\w M}_{\e})$ intersect
transversely in the neighborhood of ${\hat {\w R}}$. 
\begin{prop}
There exist a neighborhood ${\hat {\w R}}$ for the parameter $b, a, c$
and a positive constant ${\e}_{0}>0$ such that fprany ${\vert {\e}
\vert}<{\e}_{0}$ and $(b, a, c)\in {\hat {\w R}}$ the 
$W^{\st}({\w M}_{\e})$ and $W^{\un}({\w M}_{\e})$ intersect
transversely. 
\end{prop}

\subsection{Case II: Time-Periodic Perturbation}
The system (\ref{1.2b}) can be rewritten in the more general form as
follows:
\bqn
{\dot {\rx}}&=&{\rm f}({\rx})+{\e}{\rG}_{II}({\rx}, \th),\cr
{\dot {\th}}&=&1
\label{mII.1}
\eqn
We assume that ${\rG}_{II}\in C^{\infty}({\Hset}^{1}\times {\Sset}^{1},
{\Hset}^{1})$, where ${\Sset}^{1}={\Rset}/{T}$. 
This implies that the associated suspended autonomous
system on ${\Hset}^{1}\times {\Sset}^{1}$ has a smooth local flow 
${\rF}^{t}_{\e}:{\Hset}^{1}\times {\Sset}^{1}\rh {\Hset}^{1}\times
{\Sset}^{1}$, this means that ${\rF}^{t}_{\e}$ is a smooth map
defined for small ${\vert t \vert}$ which is jointly continuous in all
variables $\e, t\in {\Rset}, {\rx}\in {\Hset}^{1}, {\th}\in {\Sset}^{1}$. 

We recall that for ${\e}>0$ small, there is a unique fixed point $p_{\e}$ of 
${\rm P}^{\e}$ near $p_{0}=O=(0, 0)$, moreover $p_{\e}-p_{0}={\Or}(\e)$, i.e
there is a constant $K$ such that ${\Vert p_{\e} \Vert}<K{\e}$, 
${\forall}{\e}>0$.

We consider a neighborhood of the codimension 2
center manifold ${\w M}_{\e}$, 
in which we attempt to establish the transversal intersection of
$W^{\st}(p_{\e})\subset W^{\st}({\w M}_{\e})$ and 
$W^{\un}(p_{\e})\subset W^{\un}({\w M}_{\e})$. Based on the
assumptions and preliminary results of Lemma 3.1, we proceed
to calculate the distance $\w D$ between the perturbed manifolds
$W^{\st}(p_{\e}), W^{\un}(p_{\e})$, by calculating the ${\Or}(\e)$
components of perturbed solution curves of equation (\ref{mII.1}) from
the f\/irst variation equation. We expand solution curves in 
$W^{\st}(q_{\e})$ and
$W^{\un}(q_{\e})$, as we described in Lemma 3.1, points in
$W^{{\st}, {\un}}(p_{\e})$ are obtained by intersecting $W^{{\st},
{\un}}(q_{\e})$ with the section ${\Hset}^{1}\times \{t_{0}\}$.

Therefore, (\ref{mII.1}) possesses  
the perturbed orbit 
\[
{\rx}^{\un}_{\e}(t, t_{0})={\rx}^{\un}_{0}(t-t_{0})+{\e}
{\rx}^{\un}_{1}(t, t_{0})+{\Or}({\e}^{2})
\]
on the perturbed unstable manifold, 
where ${\rx}^{\un}_{1}(t, t_{0})$ is the solution of the f\/irst variation
equation
\[
{\dot {\rx}}^{\un}_{1}(t, t_{0})=D{\rm f}({\rx}_{0}(t-t_{0})){\rx}^{\un}_{1}(t,
t_{0})+{\rG}_{II}({\rx}_{0}(t-t_{0}), t)
\]
and a similar one lying on the perturbed stable manifold.

As in the case I, the distance function is given by:
\bq
{\w D}(t_{0})={\e}M_{II}(t_{0})+{\Or}({\e}^{2})
\label{mII.2}
\eq
with 
\bq
M_{II}(t_{0})=\int_{-\infty}^{\infty}\int_{0}^{L}\frac{{\d}{\D}}{{\d}u_t}(U)\,\big(\,{g}_{2}(t-t_{0})-a
U_{t}\big){\id}x{\id}t
\label{mII.3}
\eq
evaluated on the homoclinic orbit $U$.

For $g_{2}(t)={\G}{\cos}{\o}t$, the equation (\ref{mII.3}) takes the
form:
\bq
M_{II}(t_{0})=\int_{-\infty}^{\infty}\int_{0}^{L}\frac{{\d}{\D}}{{\d}u_t}(U)\,\big(\,{\G}{\cos}{\o}(t-t_{0})-a
U_{t}\big){\id}x{\id}t
\label{mII.4}
\eq
The Mel'nikov integral in equation (\ref{mII.4}) is convergent. We
therefore consider the improper integral as the following limit of a
sequence in time $\{ {\tau}^{{\st},
{\un}}_{n}\}=\{{\pm}2{\pi}n/{\o}\}$, $n=1,2,\ldots$
\bq
M_{II}(t_{0}, {\G}, a, {\b}, c)=\lim_{n\to
{\infty}}\int_{-\frac{2{\pi}n}{\o}}^{\frac{2{\pi}n}{\o}}\int_{0}^{L}
\frac{{\d}{\D}}{{\d}u_t}(U)\,\big(\,{\G}{\cos}{\o}(t-t_{0})-a
U_{t}\big){\id}x{\id}t
\label{mII.5}
\eq
Since $\frac{{\d}{\D}}{{\d}u_t}(U)$ and $U_{t}$ are rapidly
decreasing expressions, we are allowed to extend the integration
limits in (\ref{mII.5}) to inf\/inity and thus obtaining convergent
integral. 

We f\/ind that the Mel'nikov function is of the form
\bq
M_{II}(t_{0}, {\G}, a, {\b}, c)={\w A}_{2}{\G}{\sin}{\o}t_{0}-\ {\w I}_{2}
\label{mII.6}
\eq
where
\bqns
{\w A}_{2}{\sin}\,{\o}t_{0} & = & 
\int_{-\infty}^{\infty}\int_{0}^{L}\frac{{\d}{\D}}{{\d}u_{t}}(U(x, t))
{\cos}{\o}(t-t_{0}){\id}x{\id}t\cr\cr
{\w I}_{2} & = &
\int_{-\infty}^{\infty}\int_{0}^{L}\frac{{\d}{\D}}{{\d}u_{t}}(U(x, t))
U_{t}(x, t){\id}x{\id}t
\eqns
Setting $M_{II}(t_{0})=0$, we obtain 
\bq
t^{\pm}_{0}={\pm}\frac{1}{\o}{\sin}^{-1}\big(\,\frac{a}{\Gamma}{\chi}_{2}\big)
\quad 
{\chi}_{2}=\frac{{\w I}_{2}}{{\w A}_{2}}
\label{4.13}
\eq
\begin{lem}
For $\frac{1}{4}<c^{2}<1$, there exists an open interval $I\subset
{\Rset}$, such
that for any ${\Gamma}, a, {\b}\in I$, there are two values of $t_{0}$:
$t^{\pm}_{0}=t^{\pm}_{0}(c, {\o}, {\Gamma}, a, {\b}, {\chi}_{2})$
(cf. (\ref{4.13})) at which 
$M_{II}(t_{0}, {\mu})=0$, ${\mu}=(c, {\o}, {\Gamma}, a, {\b})$. 
Moreover, when $M_{II}(t_{0})$ is viewed as function
of $t^{\pm}_{0}$, these zeros are simple.
\end{lem}
By this lemma, eq. (\ref{mII.2}) and the implicit function theorem, we have the
following proposition:
\begin{prop}
There exists a subregion ${\hat {\w R}}_{1}$ of external parameter space 
$${\hat {\w R}}=\Big\{\,(c, {\o}, {\Gamma}, a, {\b})\,:\,\,
\frac{1}{4}< c^2 < 1,\,{\o}\in \big(\,0, 0.87\,\big),\, 
{\Gamma}, a\in \big(\, 0.1, 1\,\big),\,
{\b}\in \big(\,-{\pi}/{2}, {\pi}/{2}\,\big)\,
\Big \}$$
and a positive number ${\e}_{0}$, such that for any f\/ixed parameters 
$\{ c, {\o}, {\Gamma}, a, {\b} \}\in {\hat {\w R}}_{1}\times (\, 0,
{\e}_{0}\,)$, 
there are $t^{\pm}_{0}$ (cf. (\ref{4.13})) at which $W^{\st}(p_{\e}),
W^{\un}(p_{\e})$ intersect transversely.
\end{prop}

\section{Conclusions}
In this paper, we have used a Mel'nikov type analysis to detect the splitting
of homoclinic manifolds in a perturbed sine--Gordon equation with even
spatial periodic boundary conditions. We extended the two-dimensional phase portrait for the
pendulum into the inf\/inite dimensional phase space ${\Hset}^{1}$ 
of the integrable periodic sine--Gordon equation. 

This analysis consists f\/irst of a study of homoclinic and chaotic
dynamics on perturbed SG systems. The prerequisite for such study is that the
unperturbed soliton system, viewed as a Hamiltonian system should have a
homoclinic structure. 
In the process of proving the existence of invariant manifolds for such a 
system several mathematical tools from dynamical systems theory are utilised.
We established, via Mel'nikov theory the existence of transversal
intersection of a codimension 2 center-stable and center-unstable manifold,
of the perturbed singular point $O_{\e}$, which implies the occurrence of chaotic dynamics. In particular geometric settings, a
simple zero of the Mel'nikov function, with respect to one of its
parameters insure the intersection of stable and unstable manifolds of
a critical torus and the persistence which follows as a consequence of
this intersection. 
%\newpage
\vskip 0.6 true cm
\begin{center}
{\bf Acknowledgement}
\end{center}
\vskip 0.5 true cm
The author had stimulating discussions with T.Bountis and G.Haller. He
would also like to thank the anonymous referee for helpful comments 
and remarks.
This research was supported by the postdoctoral fellowship
No. ERBFMBICT983236 of the Commission of the European
Communities.
\vskip 0.5 true cm
%\newpage
%%%%%%%%%%%%%%%%%%%%%%

\end{document}

\appendix
\section*{Appendix. Homoclinic Solution and $grad{\Delta}(U(x, t))$.}
\label{app}
\pagestyle{myheadings}
\setcounter{equation}{0}
\setcounter{section}{1}

Let us consider the special solution of the unperturbed SG, $u=(\,0, 0\,)$. 
In this case the solutions of system (\ref{2.7}) are given by
\bq
f^{\pm}(x, t)={\exp}[{\pm}{\iu}(\frac{k(\zeta)x}{c}+{\l}(\zeta)t)]
\left ( \begin{array}{c} 1 \\ {\pm}{\iu}
\end{array} \right)
\label{B.1}
\eq
where $k(\zeta)={\zeta}+\frac{1}{16\zeta}$,
${\l}(\zeta)={\zeta}-\frac{1}{16{\zeta}}$. 
For these solutions the spectrum of $L^{(x)}$ has a double complex point at
${\nu}=\frac{1}{4}{\exp}[{\iu}{\beta}]$.
Ercolani et al. \cite{erc1} have proved that the solution (\ref{2.18}) of SG
may be obtain by a decomposition of two B\"acklund transformation from 
$u=(\,0, 0\,)$. Let $f(x, t, \zeta)=(f_1, f_2)^{\bot}(x, t, \zeta)$ denote a
general 
solution of (\ref{2.7}) at $((0, 0), \nu)$ (cf. (B.1)). Then, the solution 
$U(x, t)$ of SG given as:
\bq
U(x, t)=2{\iu}{\ln}\left(\frac{{\nu}f_{1}f^{*}_{1}+{\nu}^{*}f_{2}f^{*}_{2}}
{{\nu}^{*}f_{1}f^{*}_{1}+{\nu}f_{2}f^{*}_{2}}\right)
\label{B.2}
\eq
Substituting in (B.2) the double point ${\nu}=\frac{1}{4}exp[i{\beta}]$,
we obtain 
\bq
U(x, t)=4\Big(\frac{1}{2{\iu}}{\ln}\left(\frac{1+{\iu}{\tan}{\beta}
{\di{{f_{1}f^{*}_{1}-f_{2}f^{*}_{2}}\over {f_{1}f^{*}_{1}+f_{2}f^{*}_{2}}}}} 
{1-{\iu}{\tan}{\beta}{{\di{f_{1}f^{*}_{1}-f_{2}f^{*}_{2}}\over
{f_{1}f^{*}_{1}+f_{2}f^{*}_{2}}}}} 
\right)\Big)
\label{B.3}
\eq
making use of the identity ${\tan}^{-1}x=\frac{1}{2{\iu}}{\ln}(\frac{1+{\iu}x}{1-{\iu}x})$
and using the general solution $h(x, t, \nu)$ where $h_{1}$, $h_{2}$ are
evaluated at ${\nu}=\frac{1}{4}{\exp}[{\iu}{\beta}]$, it follows after some algebra
that 
$$U(x, t)=4{\tan}^{-1}\Big (
{\tan}\,{\b}\,{\cos}\,({x}{c}\,{\cos}\,{\b})\,{\sech}\,(t\,{\sin}\,{\b}) \Big)
$$
In order to compute the second component of Mel'nikov vector (cf.
(\ref{4.11})), we need the gradient of teh Floquet discriminant evaluated on
the homoclinic orbit $U(x, t)$. 

Let $m(x, x^{\prime}, t, \zeta, O)=[f_{ij}]$ denote the fundamental matrix of
(\ref{2.7}) at $(\zeta, O)$, ${\psi}$ the general solution of Lax pair
system and $U(x, t)$ the homoclinic orbit. 
Let $M(x, x^{\prime}, t, \zeta, U)$ denote the
fundamental matrix of (\ref{2.7}) at $(\zeta, U(x, t))$. Then, for
${\zeta}^{2}\neq {\nu}^{2}$, $M$, $m$ are related by
\bq
M(x, t, \zeta, U)=G(x, t; \zeta, \nu)m(x, x^{\prime},t, \zeta, O)
G^{-1}(x, t; \zeta, \nu)
\label{B.4}
\eq
where the matrix $G$ is given by
\bq
G(x, t; \zeta, \nu)=\left ( \begin{array}{cc} 
-{\nu}\frac{{\psi}_{1}}{{\psi}_{2}}  &  \zeta \\ -\zeta  &
{\nu}\frac{{\psi}_{2}}{{\psi}_{1}} \end{array} \right)
\label{B.5}
\eq
Using the fact that the homoclinic orbit (\ref{2.18}) is generated
from $O$ with a composition of two B\"acklund transformation (at ${\zeta}={\nu}$ and
${\zeta}=-{\nu}^{*}$). It follows after some manipulation that the fundamental
matrix $M$ evaluated on $U$ given by
\bq
M(x, t, \zeta, U)=\left ( \begin{array}{cc} 
M_{11}  &  M_{12} \\ M_{21} & M_{22} \end{array} \right)
\label{B.6}
\eq
where
\bqn
M_{11} & := & \frac{\iu}{2{\sin}2{\beta}}\left[ \big(e^{-2{\iu}{\b}}f_{22}-
e^{2{\iu}{\b}}f_{11}\big)+\Big(
f_{12}\frac{{\psi}_{2}}{{\psi}_{1}}-f_{21}\frac{{\psi}_{1}}
{{\psi}_{2}}\Big)\,\right]\cr\cr 
M_{12} & := & \frac{-{\iu}}{2{\sin}2{\beta}}\left[
\frac{{\psi}_{2}}{{\psi}_{1}}\big(f_{11}-f_{22}\big)-e^{2{\iu}{\b}}f_{12}\big(\frac{{\psi}_{2}}{{\psi}_{1}}\big)^{2}
+e^{-2{\iu}{\b}}f_{21}\,\right]\cr\cr
M_{21} & := & \frac{-{\iu}}{2{\sin}2{\beta}}\left[ \frac{{\psi}_{1}}{{\psi}_{2}}\big(f_{11}-f_{22}\big)+e^{2{\iu}{\b}}f_{21}\big(\frac{{\psi}_{1}}{{\psi}_{2}}\big)^{2}-e^{-2{\iu}{\b}}f_{12}\,\right]\cr\cr
M_{22} & := & \frac{\iu}{2{\sin}2{\beta}}\left[ \big(e^{-2{\iu}{\b}}f_{11}-
e^{2{\iu}{\b}}f_{22}\big)-\Big(f_{12}\frac{{\psi}_{2}}{{\psi}_{1}}-f_{21}
\frac{{\psi}_{1}}{{\psi}_{2}}\Big)\,\right]
\label{B.6a}
\eqn
We prove that the gradient of the Floquet discriminant ${\D}({\bf u},
\zeta)$ is given in (\ref{2.17}). 
We rewrite the linear operator $L^{(x)}({\bf u}, \zeta)$
(cf. (\ref{2.6})) as follows
\bq
L^{(x)}({\bf u}, \zeta):=-c{\rm J}D_{x}+\big(\,{\rm A}+
\frac{{\rm B}^{2}}{\zeta}-{\zeta}\,\big)I
\label{D1}
\eq
where 
\[
{\rm J}=\left ( \begin{array}{cc} 0 & 1 \\ -1 & 0 \end{array}
\right),\quad
{\em A}=\frac{\iu}{4}(cu_x+u_t)\left ( \begin{array}{cc}
0 & 1 \\ 1 & 0 \end{array} \right)
\]
and 
\[
{\rm B}=\frac{\iu}{4}\left (
\begin{array}{cc} e^{{\iu}u/2} & 0 \\ 0 & e^{-{\iu}u/2} \end{array}
\right)
\]
Let $M$ be the fundamental matrix to the system 
\bq
L^{(x)}({\bf u}, \zeta)M=0,\quad M(x=0)=I
\label{D2}
\eq
or equivalent 
\[
D_{x}M=\frac{1}{c}{\rm J}\big(\,{\zeta}-\big(\,{\rm A}+\frac{{\rm
B}^{2}}{\zeta}\,\big)\,\big)M
\]
Variations of ${\bf u}$ leads to the variational equation for the
variation of $M$ at fixed $\zeta$:
\bqns
D_{x}{\d}M & = & \frac{1}{c}{\rm J}\Big(\,{\zeta}-\Big(\,{\rm A}+
\frac{{\rm B}^{2}}{\zeta}\,\Big)\,\Big){\d}M-\frac{\rm J}{c}
\Big(\,{\d}{\rm A}+\frac{2{\rm B}{\d}{\rm B}}{\zeta}\,\Big)M\cr\cr
{\d}M(x, {\zeta}, {\bf u}) & = &
-\frac{1}{c}M(x)\di{\int_{0}^{x}}M^{-1}(y){\rm J}\Big[\frac{\iu}{4}\big[{\d}u_{t}+c{\d}u_{x}\big]\left ( \begin{array}{cc}
0 & 1 \\ 1 & 0 \end{array} \right)\cr\cr 
& - &\frac{{\iu}{\d}u}{16{\zeta}}\left (
\begin{array}{cc} e^{{\iu}u} & 0 \\ 0 & -e^{-{\iu}u} \end{array}
\right)\,\Big]M(y){\id}y\cr\cr
{\d}{\D}(L, {\zeta}, {\bf u}) & = &
-\frac{1}{c}{\trace}\bigg[\,M(L)\di{\int_{0}^{L}}M^{-1}(y){\rm J}\Big[\frac{\iu}{4}\big[{\d}u_{t}+c{\d}u_{x}\big]\left ( \begin{array}{cc}
0 & 1 \\ 1 & 0 \end{array} \right)\cr\cr
& - & \frac{{\iu}{\d}u}{16{\zeta}}\left (
\begin{array}{cc} e^{{\iu}u} & 0 \\ 0 & -e^{-{\iu}u} \end{array}
\right)\,\Big]M(y){\id}y\,\bigg]
\eqns
An explicit calculation yields the equation (\ref{2.17}).

Using the fact that the elements of the
matrix $M$ are $L$-periodic functions in $x$ the second equation in 
(\ref{2.17}) takes the form
\bq
\frac{{\d}{\D}}{{\d}u_t}(U(x, t))=-\frac{\iu}{4c}{\trace}\bigg [ M^{-1}(x)
\left ( \begin{array}{cc} 1 & 0 \\ 0 & -1 \end{array} \right) 
M(x+L)\bigg ]
\label{B.7}
\eq
Insertion of (\ref{B.6a}) into (\ref{B.7}) leads, after simple algebra,
to the following expression for ${\d}{\D}/{\d}u_{t}$
\bqn
\frac{{\d}{\D}}{{\d}u_{t}}(U(x, t)) & = & -\frac{\iu}{4c}
\bigg[\, \Big(\,M_{22}(x)M_{11}(x+L)-M_{11}(x)M_{22}(x+L)\,\Big)\cr\cr
& + & \Big(\,M_{12}(x)M_{21}(x+L)-M_{21}(x)M_{12}(x+L)\,\Big)\,\bigg]
\label{B.7a}
\eqn

\end{document}